\documentclass[prd,aps,a4paper,nofootinbib,eqsecnum,twocolumn]{revtex4}  

\newif\ifusesec
\usesectrue  
   
\usepackage{graphicx} 
\usepackage{mathrsfs}
\usepackage{amsmath,amsfonts,amssymb}
\usepackage{multirow}

\newcommand{\beq}{\begin{equation}}
\newcommand{\eeq}{\end{equation}}
\newcommand{\bea}{\begin{eqnarray}}
\newcommand{\eea}{\end{eqnarray}}

\begin{document}

\title{Nariai spacetime: orbits, scalar self force and 
Poynting-Robertson-like external force}

\author{Donato \surname{Bini}$^{1}$, Giampiero Esposito$^{2,3}$}

\affiliation{$^1$Istituto per le Applicazioni del Calcolo ``M. Picone'', 
CNR, I-00185 Rome, Italy\\
}
\affiliation{$^2$Dipartimento di Fisica ``Ettore Pancini'', \\
Complesso Universitario di Monte S. Angelo,
Via Cintia Edificio 6, 80126 Napoli, Italy\\
} 
\affiliation{$^3$Istituto Nazionale di Fisica Nucleare, Sezione di Napoli, \\
Complesso Universitario di Monte S. Angelo,
Via Cintia Edificio 6, 80126 Napoli, Italy}

\begin{abstract}
After studying properties of the Nariai solution, including its geodesics, in spherical and
de Sitter coordinates, two kinds of accelerated motion are investigated in detail:
either observers at rest with respect to the coordinates, or observers
in radial motion.
Next, massless scalar perturbations of Nariai spacetime in absence
of sources are worked out, and an explicit example out of the 
black hole context of analytic self-force
calculation is obtained. Last, self-force effects are studied as well, 
together with some variant of the type of Poynting-Robertson external force,
and also building a test electromagnetic field and a test gravitational
field in Nariai spacetime geometry.
\end{abstract}

\date{\today}

\maketitle

\section{Introduction}

In general relativity, an important exact solution of Einstein's
field equations in the presence of cosmological constant was 
found by Nariai \cite{Nariai1,Nariai2} in 1951.
This spacetime can be described as the topological product 
$dS_2 \times S_2$ (with decomposable metric), where
$dS_2$ is a 2-dimensional de-Sitter (dS) spacetime and $S_2$ 
is a 2-sphere of constant radius. 
Originally the metric was written in spherical-like coordinates in the form
\begin{equation}
ds^2=-dt^2+\cosh^2 t\, dr^2 +ds^2_{(\theta,\phi)},
\label{(1.1)}
\end{equation}
where
\begin{equation}
ds^2_{(\theta,\phi)}=d\theta^2+\sin^2\theta d\phi^{2}.
\label{(1.2)}
\end{equation}
Its $(t,r)$ section,
\begin{equation}
ds^2_{(t,r)}=-dt^2+\cosh^2 t\, dr^{2},
\label{(1.3)}
\end{equation}
represents a pseudo-sphere, for which it is known that there exist various coordinate 
representations. For example, \eqref{(1.1)} can be cast in the dS-like form
\begin{equation}
ds^2=ds^2_{(T,\rho)}+ds^2_{(\theta,\phi)},
\label{(1.4)}
\end{equation}
with
\begin{equation}
ds^2_{(T,\rho)}= -(1-\rho^2)dT^2+\frac{d\rho^2}{(1-\rho^2)},
\label{(1.5)}
\end{equation}
via the nontrivial coordinate map
\begin{equation}
\rho=\cosh t \cos r ,\qquad T={\rm arctanh}\left(\frac{\sin r}{\tanh t}\right).
\label{(1.6)}
\end{equation}
Here the $(T,\rho)$ section also represents a pseudo-sphere, but the 
\lq\lq local" properties (not the \lq\lq global" ones) of the metric 
are very different in the two coordinate patches.
For instance, in the first case the metric is time-dependent whereas in 
the second case it is static. Thus, kinematical problems 
(motions) are formally different when studied in the two cases, but clearly 
traceable one to the other via the transformation \eqref{(1.6)}.
Furthermore, observers adapted to the two sets of coordinates, resulting in relative 
motion one with respect to the other, can play special roles.
From a practical point of view, it is a lucky circumstance that here one has 
the possibility to switch from one form of the metric to the other, and 
look for simplifying situations, as we will systematically do in the following.
A final remark concerns the range of variability of the coordinates. 
Besides the standard spherical coordinate ranges $\theta\in[0,\pi]$, 
$\phi\in[0,2\pi)$, we will assume below $t,T\in {\mathbb R}$, $r\ge 0$,  
whereas  $\rho\in (-1,1)$, i.e., $\rho$ is not a polar coordinate radius. 

Nariai spacetime has several relevant global properties \cite{Ginsparg:1982rs}: 
constant curvature, geodesic completeness, global hyperbolicity, algebraic 
speciality of Petrov type D, etc. Furthermore, as is evident when using the 
form \eqref{(1.4)} of the metric, it is 
related to the dS-Schwarzschild black-hole-like solution during its thermodynamical 
equilibrium phase (which is but a physical property). More specifically, 
following Ref. \cite{Ginsparg:1982rs}, it is
generated if the black hole event horizon of the dS-Schwarzschild solution approaches the
cosmological horizon (through an appropriate limiting procedure).  

Here, after reviewing some curvature aspects, and relating the sectional 
curvatures of a timelike 2-section and of a spacelike 2-section to the 
spacetime curvature itself by analyzing the Kretschmann invariant properties, we study both 
geodesic motion (arriving at a complete, explicit analytic 
integration of the orbits in both coordinate systems) and accelerated one. 
In the latter case, the most natural types of 
accelerated motion are associated with observers at rest with respect to the coordinates 
or in radial motion with respect to them. We discuss both families of fiducial observers.  

Besides this, we investigate  
scalar field perturbations (occurring in the scalar gravitational self-force 
problem), both in absence of sources and in the non-minimal coupling situation within 
the conformally invariant case, as well as with inclusion of scalar charge sources, 
always exploring separability of the associated equations in both types of coordinates 
as well as the possibility to Fourier-decompose the field with respect 
to some coordinate (typically time or radius). 
Furthermore, we study scalar self-force effects, together with other acceleration 
mechanisms related to what is known as the Poynting-Robertson(-like) approach.
Last, a test electromagnetic field in Nariai geometry is built in Sec. X,
and the gravitational counterpart is addressed in Sec. XI.

\section{Properties of the Nariai solution in spherical-like coordinates}

Let us start our investigation by using the Nariai spacetime  \cite{Nariai1,Nariai2} 
metric in the form \eqref{(1.1)}, which seems to be less studied.
Such a  metric, that we have written in dimensionless 
spherical-like coordinates $(t,r,\theta,\phi)$  
\footnote{Dimensionful coordinates $[T,R]$ can be restored, for example, by 
using $\Lambda\sim 1/L^2$ as an overall length scale, that is
$$
ds^2=-dT^2+ \cosh^2 (T\sqrt{\Lambda})\, dR^2 
+\frac{1}{\Lambda}[d\theta^2+\sin^2\theta d\phi^2].
$$
This metric solves Einstein's equation $G_{\mu\nu}+\Lambda g_{\mu\nu}=0$, 
i.e., for generic value of $\Lambda$. Furthermore $R=+4\Lambda$.}
is an exact solution of Einstein's field equation with 
cosmological constant\footnote{Notice that the sign of $\Lambda$ depends 
on the Ricci tensor definition; here the definition is 
$R_{\alpha\beta}=R^\mu{}_{\alpha\mu\beta}$ and the metric signature 
is $-+++$.} $\Lambda=+1$, i.e., with
\begin{equation}
G_{\alpha\beta}=-g_{\alpha\beta} ,
\label{(2.1)}
\end{equation}
where $G_{\alpha\beta}$ denotes the Einstein's tensor.
In spite of its apparent formal simplicity it has several important properties \cite{Kroon,Beyer}
already mentioned in Sec. I:
it is geodesically complete,  globally hyperbolic and algebraically special of Petrov type D.
Furthermore, as is discussed in Ref. \cite{Kroon}, for Nariai spacetime the 
Penrose conformal boundary cannot be defined.
 
The Nariai metric is $2+2$-decomposable according to Eqs. 
\eqref{(1.1)}-\eqref{(1.3)}, where the pseudosphere part  
can be used to visualize the modifications to the light cones as 
time evolves. For instance, looking at the family of points
$(r_0,t_0)$ with $r_0$ fixed and $t_0$ taken as a parameter, 
the ($t-r$ section of the) light-cone equation becomes
\begin{equation}
t-t_0=\pm (\cosh  t_0) (r-r_0).
\label{(2.2)}
\end{equation}
This relation leads to the $45$ degree opening angle at $t_0=0$, and 
then the angle gets restricted continuously, vanishing in the 
limit $t_0\to \infty$. Differently, with $t_0$ fixed and $r_0$ taken 
as a parameter, the light-cone structure is fixed itself, i.e., the 
opening angle does not depend on $r_0$.

The condition of being locally $dS_2 \times S_2$ and with a decomposable 
metric implies that Nariai's spacetime can be derived from an 
embedding into a six-dimensional flat 
spacetime \cite{Ortaggio,Wardell,Casals} 
which is locally $H_3\oplus E_3$, i.e., 
a hyperbolic three-dimensional metric and an Euclidean three-dimensional one,
\begin{eqnarray}
ds^2&=&-(dX^0)^2+(dX^1)^2+(dX^2)^2
\nonumber\\
&+& (dX^3)^2+ (dX^4)^2+ (dX^5)^2 ,
\label{(2.3)}
\end{eqnarray}  
where
\begin{eqnarray}
X^0&=&  \sinh t  ,\quad
X^1=  \cosh t \cos r ,\quad
X^2=  \cosh t \sin r
\nonumber\\
X^3&=&  \sin\theta \cos \phi,\quad
X^4=  \sin\theta \sin \phi,\quad
X^5=  \cos\theta.\qquad
\label{(2.4)}
\end{eqnarray}
Indeed, 
\begin{equation}
-(X^0)^2+(X^1)^2+(X^2)^2=1 ,
\label{(2.5)}
\end{equation}
and
\begin{equation}
(X^3)^2+(X^4)^2+(X^5)^2=1 ,
\label{(2.6)}
\end{equation}
are the equations of the hyperboloid and the sphere, respectively.
An open question is that of deriving Nariai spacetime directly from 
a genuine embedding into a five-dimensional Euclidean space. 

The metric \eqref{(1.1)} can be viewed as a perturbation of the 
four-dimensional spherical-like background (B) metric, in the sense 
that, from the identity $\cosh^{2}t=1+\sinh^{2}t$,
one gets the equivalent form
\begin{eqnarray}
ds^2&=&-dt^2+dr^2 +d\theta^2+\sin^2\theta d\phi^2+\sinh^2 t\, dr^2
\nonumber\\
&=& (g^{\rm B}_{\alpha\beta}+h_{\alpha\beta})dx^\alpha dx^\beta,
\label{(2.7)}
\end{eqnarray}
with
\begin{equation}
g^{\rm B}_{\alpha\beta}dx^\alpha dx^\beta=-dt^2+dr^2 +d\theta^2+\sin^2\theta d\phi^2,
\label{(2.8)}
\end{equation}
and
\begin{equation}
h_{\alpha\beta}dx^\alpha dx^\beta=h_{rr}dr^2=\sinh^2 t\, dr^2.
\label{(2.9)}
\end{equation}
At $t=0$ the perturbation vanishes but the metric is not flat anyway; 
it would be flat if the angular part were $r^2(d\theta^2+\sin^2\theta d\phi^2)$. 

Let us introduce the 1-form\footnote{In abstract notation the fully 
covariant form of a tensor is denoted by the symbol $\flat$.} 
${\mathcal N}^\flat$ and its associated vector ${\mathcal N}$,
\begin{equation}
{\mathcal N}^\flat =\sinh t\, dr\,,\qquad {\mathcal N}
=\frac{\sinh t}{\cosh^2 t}\partial_r=\tanh t e_1 ,
\label{(2.10)}
\end{equation}
with $e_1=\frac{1}{\cosh t}\partial_r$ the unit vector of the radial direction.
Notably, for $t<0$ the vector ${\mathcal N}$ points radially inwards, 
whereas for $t>0$ the vector ${\mathcal N}$ points radially outwards.
At $t=0$, instead, the vector ${\mathcal N}$ vanishes, i.e., it is not defined.
Therefore, $h^\flat={\mathcal N}^\flat \otimes {\mathcal N}^\flat $, 
i.e., ${\mathcal N}$ is responsible for spacetime deviations from the 
spherical background behavior $g^{\rm B}_{\alpha\beta}dx^\alpha dx^\beta$.
In other words, as soon as $t$ increases from the initial value $t=0$ 
(where the metric is given by  Eq. \eqref{(2.8)}), 
the spacetime gets warped along the radial direction. 

Even though it seems very natural, the split \eqref{(2.7)} is not the only possible one. 
Indeed, one can also decompose the metric as
\begin{equation}
ds^2=ds^2_{\rm flat}+\sinh^2 t\, dr^2+(1-r^2)(d\theta^2+\sin^2\theta d\phi^2)
\end{equation}
with 
\begin{equation}
ds^2_{\rm flat}=-dt^2+dr^2+r^2(d\theta^2+\sin^2\theta d\phi^2).
\end{equation}
In this case, deviations from flat spacetime become manifest as 
$r$ is far from $\pm 1$ and $t$ increases from $0$.

\subsection{Curvature invariants}

The Riemann tensor of the Nariai spacetime has the following independent 
nonvanishing components: 
\begin{equation}
R_{\theta\phi\theta\phi}= \sin^2 \theta ,\qquad  R_{trtr}= -\cosh^2 t ,
\label{(2.11)}
\end{equation}
corresponding to the curvatures of the $\theta-\phi$ and $t-r$ sections. 
For $t=0$ these components reduce to 
\begin{equation}
R_{\theta\phi\theta\phi}^{\rm B}= \sin^2 \theta ,\qquad  R_{trtr}^{\rm B}= -1 .
\label{(2.12)}
\end{equation}

The Kretschmann curvature invariant turns out to be constant
\begin{equation}
{\mathcal K}=R^{\alpha\beta\gamma\delta}R_{\alpha\beta\gamma\delta}=8 ,
\label{(2.13)}
\end{equation}
as for a spacetime which is nowhere singular. Unfortunately, the above result 
\eqref{(2.13)} implies that the spacetime is nowhere close to a flat spacetime, that 
is a strong limitation for all perturbative approaches taking 
advantage of an \lq\lq almost flat" or weak field situation.   
Also the (right) dual of the Riemann tensor is nonvanishing, with the 
single independent component 
\begin{equation}
R^*_{tr\theta\phi}= -\frac{1}{\cosh t \sin \theta} .
\label{(2.14)}
\end{equation}
Differently, because of the underlying spherical symmetry, 
the invariant ${\mathcal H}=R^{\alpha\beta\gamma\delta}
R^*_{\alpha\beta\gamma\delta}=0$ vanishes identically.

Let us also recall that the 2-sphere with metric $ds^2_{(\theta,\phi)}$ has 
its Kretschmann curvature invariant given by
\begin{equation}
{\mathcal K}^{\rm 2d\, sph}=4\,,
\label{(2.15)}
\end{equation}
whereas the two-dimensional pseudo-sphere $ds^2_{(t,r)}$ has its 
Kretschmann curvature invariant given by
\begin{equation}
{\mathcal K}^{\rm 2d\, p-sph}=4\,,
\label{(2.16)}
\end{equation}
so that ${\mathcal K}={\mathcal K}^{\rm 2d\, sph}+{\mathcal K}^{\rm 2d\, p-sph}$, 
a sort of variant (with the sum instead of the product) of the famous Gauss's \lq\lq theorema egregium."
For completeness, the  Kretschmann curvature at $t=0$ is ${\mathcal K}^B=4$, 
as expected because the $t-r$ part of the metric in this case is flat.

Last, let us note that for Nariai spacetime the Riemann tensor itself is 
covariantly constant, i.e.,
\begin{equation}
R_{\alpha\beta\gamma\delta ; \mu}=0\,,
\label{(2.17)}
\end{equation}
i.e.,  it is parallely transported along all possible directions.

\subsection{Newman-Penrose formalism}

A natural orthonormal frame for Nariai spacetime \eqref{(1.1)} is the following:
\begin{eqnarray}
e_0&=& \frac{\partial}{\partial t},
\qquad e_1
=\frac{1}{\cosh t} \frac{\partial}{\partial r},
\nonumber\\
e_2&=& \frac{\partial}{\partial \theta},
\qquad e_3=\frac{1}{\sin\theta} \frac{\partial}{\partial \phi},
\label{(2.18)}
\end{eqnarray}
which can be used to form a standard Newman-Penrose (NP) frame
\begin{eqnarray}
l&=&\frac{1}{\sqrt{2}}(e_0+e_1)\,,\qquad n=\frac{1}{\sqrt{2}}(e_0-e_1),
\nonumber\\
m&=&\frac{1}{\sqrt{2}}(e_2+ie_3).
\label{(2.19)}
\end{eqnarray}
The nonvanishing spin coefficients reduce to
\begin{eqnarray}
\gamma&=&-\epsilon
=-\frac{\sqrt{2}}{4}\tanh t
,\nonumber\\
\beta&=&-\alpha 
=\frac{\sqrt{2}}{4}\cot\theta.
\label{(2.20)}
\end{eqnarray}
The only nonvanishing Weyl scalar is $\psi_2$ and it is a constant:
\begin{equation}
\psi_2=-\frac{1}{3},
\label{(2.21)}
\end{equation}
from which the algebraically special nature of this metric (type D) follows easily.
Noticeably, this NP frame  is a principal one and therefore 
$l$ and $n$ are the two repeated Principal Null Directions (PNDs) of this spacetime.

A slight modification consisting in a rescaling of the real 
null vectors $l$ and $n$ of the NP tetrad
\begin{equation}
\label{new_NP_tetrad}
l\to \hat l =\frac{l}{\cosh t}, \qquad n\to \hat n =n \cosh t ,
\end{equation}
(with $m$ not modified) makes it possible to simplify the spin coefficients so 
that $\hat \epsilon= 0$. In fact the new nonvanishing spin coefficients are now
\begin{eqnarray}
\hat \gamma = -\frac{\sqrt{2}\sinh t }{2}\,,\qquad \hat \beta &=&-\hat \alpha  
=\frac{\sqrt{2}}{4}\cot\theta , \qquad
\label{(2.23)}
\end{eqnarray}
and the Weyl scalars remain invariant.
With $\epsilon=0$ the NP frame \eqref{new_NP_tetrad} besides being a principal 
frame is also a Kinnersley-like one, analogue 
to the Kinnersley frame in black hole spacetimes \cite{Kinnersley:1969zza}.

\subsection{Geodesics}

The spherical-like coordinates are adapted to the two spacelike Killing 
vector fields
\begin{equation}
\frac{\partial}{\partial r},
\qquad 
\frac{\partial}{\partial \phi}.
\label{(2.24)}
\end{equation} 
The existence of these Killing symmetries (see Appendix \ref{appB} for a 
detailed discussion) leads to separability of  
geodesic equations. On denoting their parametric equations 
by $x^\alpha=x^\alpha(\tau)$, one has
\begin{eqnarray}
\frac{dt}{d\tau}&=&  \sqrt{C_2+\frac{C_1^2}{\cosh^2 t }},
\nonumber\\
\frac{dr}{d\tau}&=& \frac{C_1}{\cosh^2 t }\,,\nonumber\\
\frac{d\theta}{d\tau}&=& \pm \sqrt{C_3-\frac{L^2}{\sin^2\theta}},
\nonumber\\
\frac{d\phi}{d\tau}&=&\frac{L}{\sin^2 \theta } ,
\label{(2.25)}
\end{eqnarray}
where the $C_i$ and $L$ are constant and we have chosen 
$dt/d\tau>0$ so as to obtain future-oriented orbits (with $C_1>0$ momentarily here, for simplicity 
\footnote{We will relax this condition later when discussing applications.}).
Interestingly, these equations are pairwise coupled, i.e., the coupled 
variables are $(t,r)$ (corresponding to geodesic motion on the 
pseudosphere) and $(\theta, \phi)$ (corresponding to geodesic 
motion on the $2$-sphere), with no mutual intersections. Evidently, this property 
is inherited from the orthogonally transitive form of the metric.
For instance, by assuming $L=0$ and $C_3=0$ one has $\theta(\tau)
=\theta(0)$, $\phi(\tau)=\phi(0)$ and one is left with the temporal 
and radial equations only.  
The normalization condition for these orbits: $U\cdot U=-\epsilon$, 
with $U^\alpha=\frac{dx^\alpha}{d\tau}$ and $\varepsilon=[1,0,-1]$ for timelike, 
null, spacelike orbits, respectively, reads as
\begin{equation}
\label{normaliz}
-\left(\frac{dt}{d\tau}\right)^2+\left(\frac{d\theta}{d\tau}\right)^2
+\frac{C_1^2}{\cosh^2 t }+\frac{L^2}{\sin^2 \theta }+\varepsilon=0,
\end{equation}
where only the Killing relations have been used. Inserting in Eq. 
\eqref{normaliz} the full set of Eqs. \eqref{(2.25)} one finds
\begin{equation}
C_2=C_3+\varepsilon,
\label{(2.27)}
\end{equation}
meaning that, once a causality condition is imposed, the integration 
constants $C_2$ and $C_3$ are no longer freely specifiable. An example
of numerical integration of timelike geodesics ($\varepsilon=1$) is
given in Fig. 1.

\begin{figure*}
\[
\begin{array}{cc}
\includegraphics[scale=0.3]{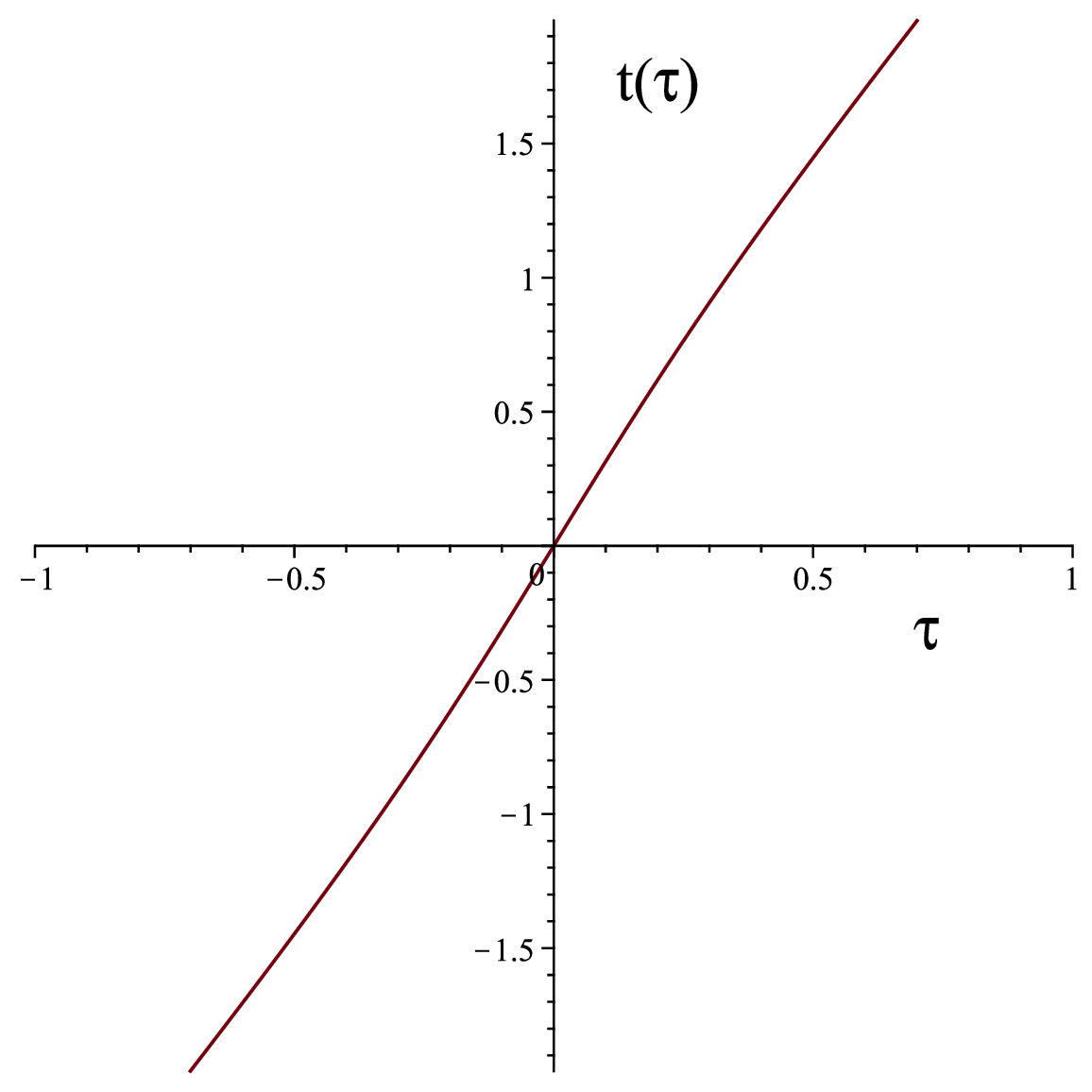}& \includegraphics[scale=0.3]{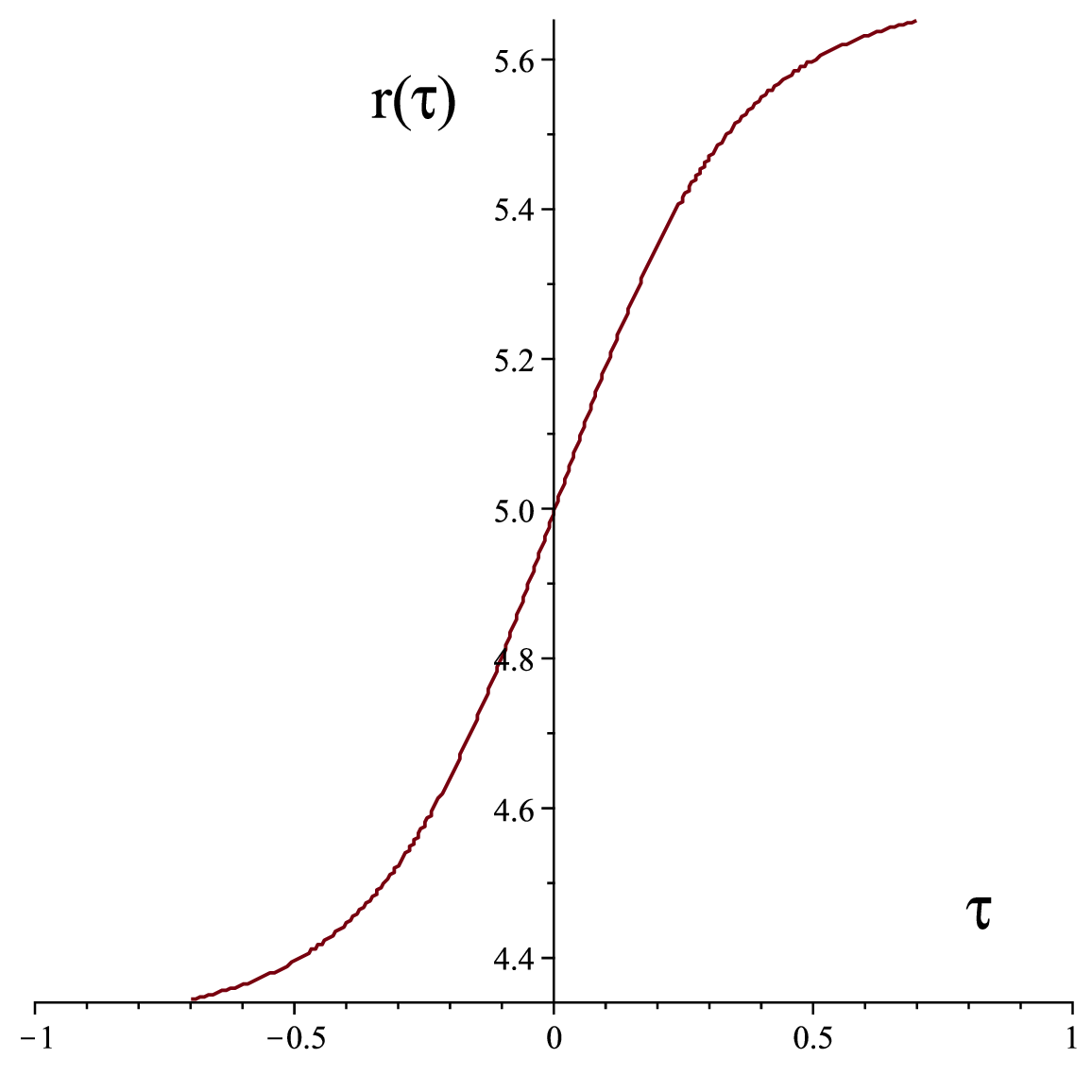}\cr
\includegraphics[scale=0.3]{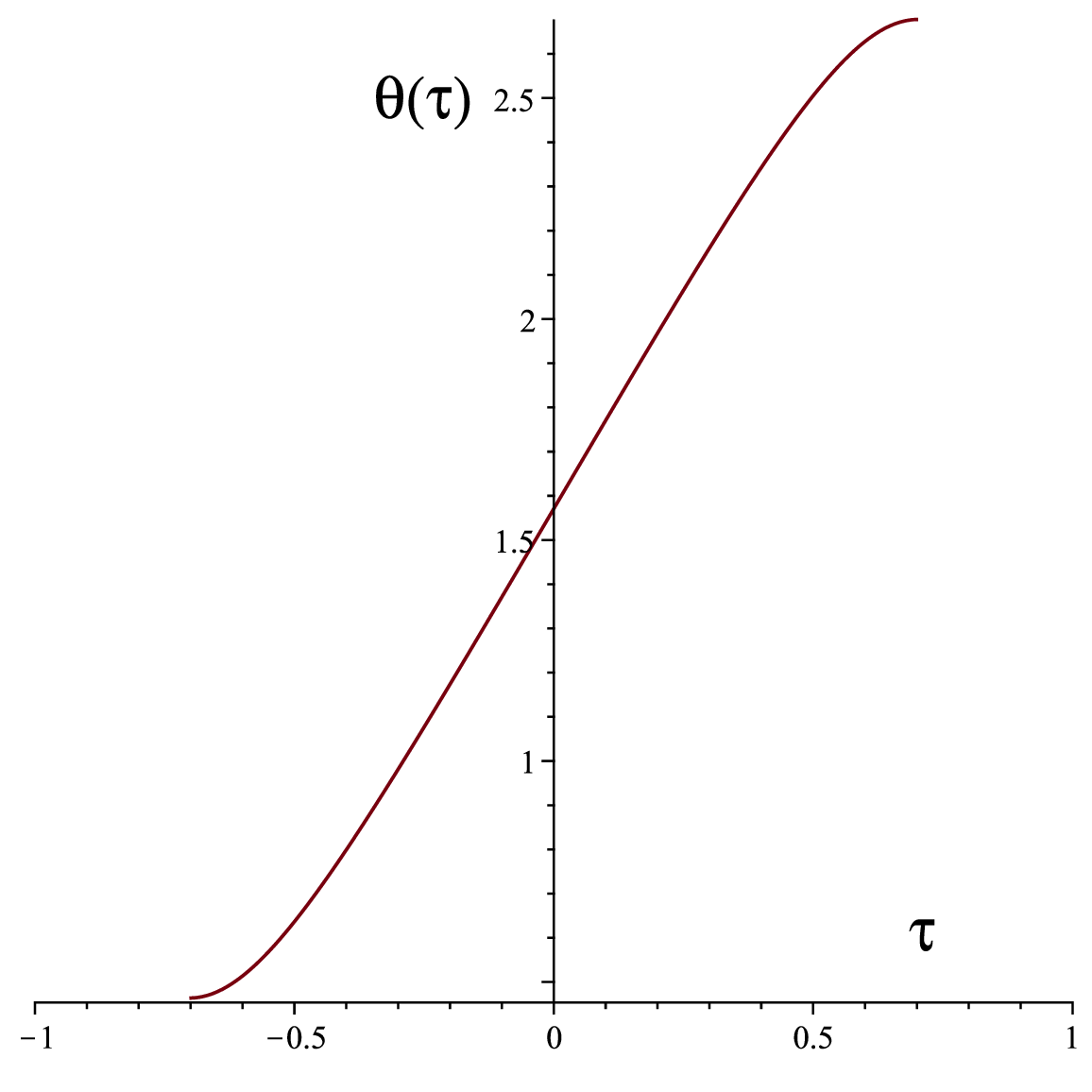}& \includegraphics[scale=0.3]{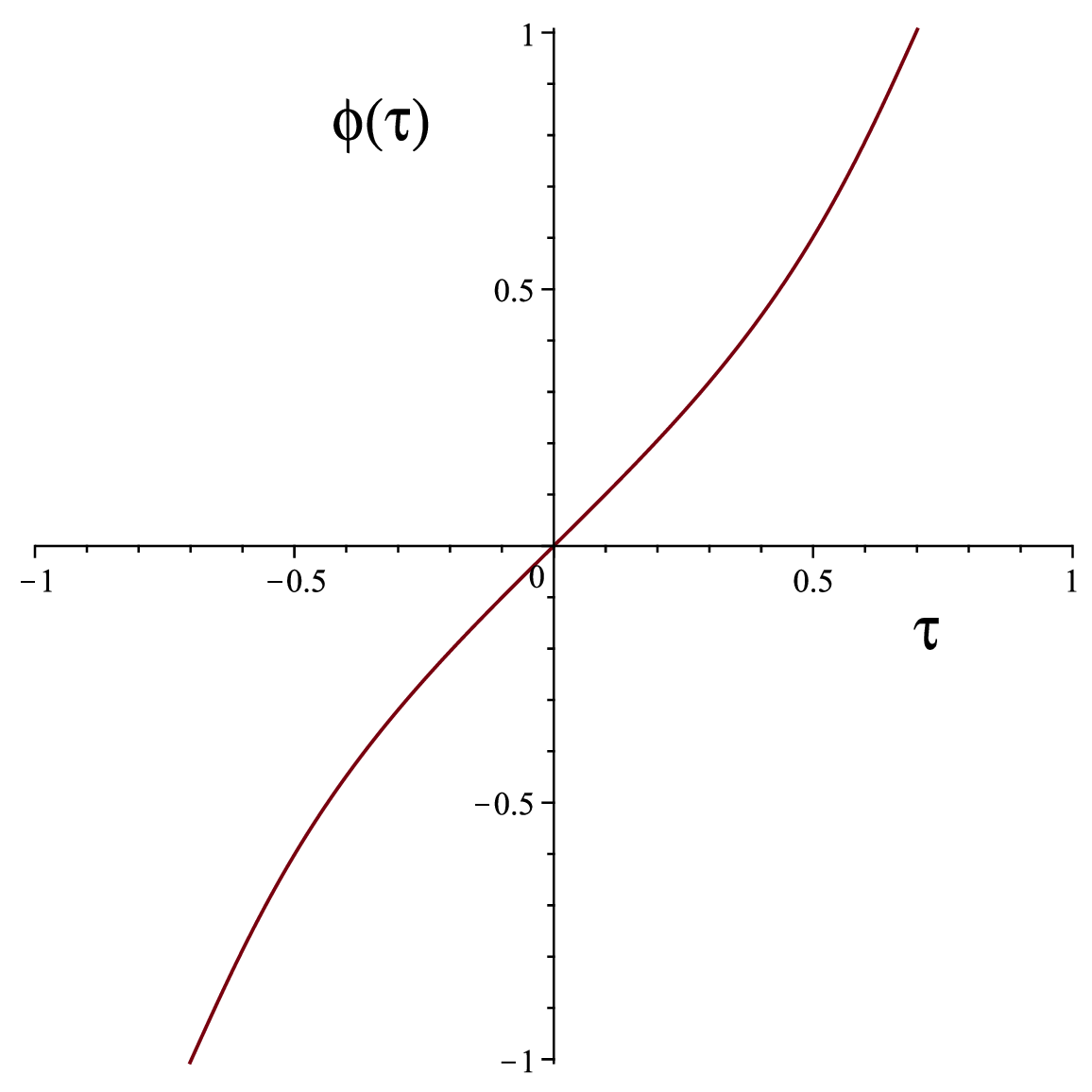}\cr
\end{array}
\]
\caption{
\label{fig:1} 
Example of numerical integration 
of timelike geodesics ($\epsilon=1$) in Nariai spacetime with parameters
$C_1=2$, $L=1$, $C_3=5$,
and initial conditions $\phi(0) = 0$, $r(0) = 5$, $t(0) = 0$, $\theta(0) = \frac{\pi}{2}$, implying
$m=\frac{3}{2}$, $\theta_*\approx 26.57{}^\circ$ and $r_\infty-r_*\approx -0.1007$. 
The maximum/minimum value of $\tau$ which follows from numerical study is 
$\tau_{\rm max/min}=\pm 0.7025$, in correspondence of which one has (within 
the numerical accuracy) reached the asymptotic values for coordinate components of the orbit: 
$t(\tau_{\rm max})\approx 1.9635$,  $r(\tau_{\rm max})\approx 5.6536$, 
$\theta(\tau_{\rm max})\approx 2.6779$, $\phi(\tau_{\rm max})\approx 1.0094$.
The value $r(\tau_{\rm max})\approx 5.6536$ fixes then $r_*\approx 5.7542$. 
Roughly speaking, these plots show the effect of the gravitational field associated 
with Nariai spacetime, which   can be taken to be that of a spacetime cosmological model. 
In fact, starting at rest at some spacetime point, say $P_0$, the (radial) expansion  
of spacetime forces all (neutral) particles to move so as to increase their radial 
position, with the subsequent, associated effect of increasing/decreasing (depending on 
the signs of the various orbital parameters) also the other coordinates. As soon as an 
asymptotic radial position is reached, asymptotic values for the other coordinates 
are reached as well.
}
\end{figure*}

\begin{figure}
\includegraphics[scale=0.3]{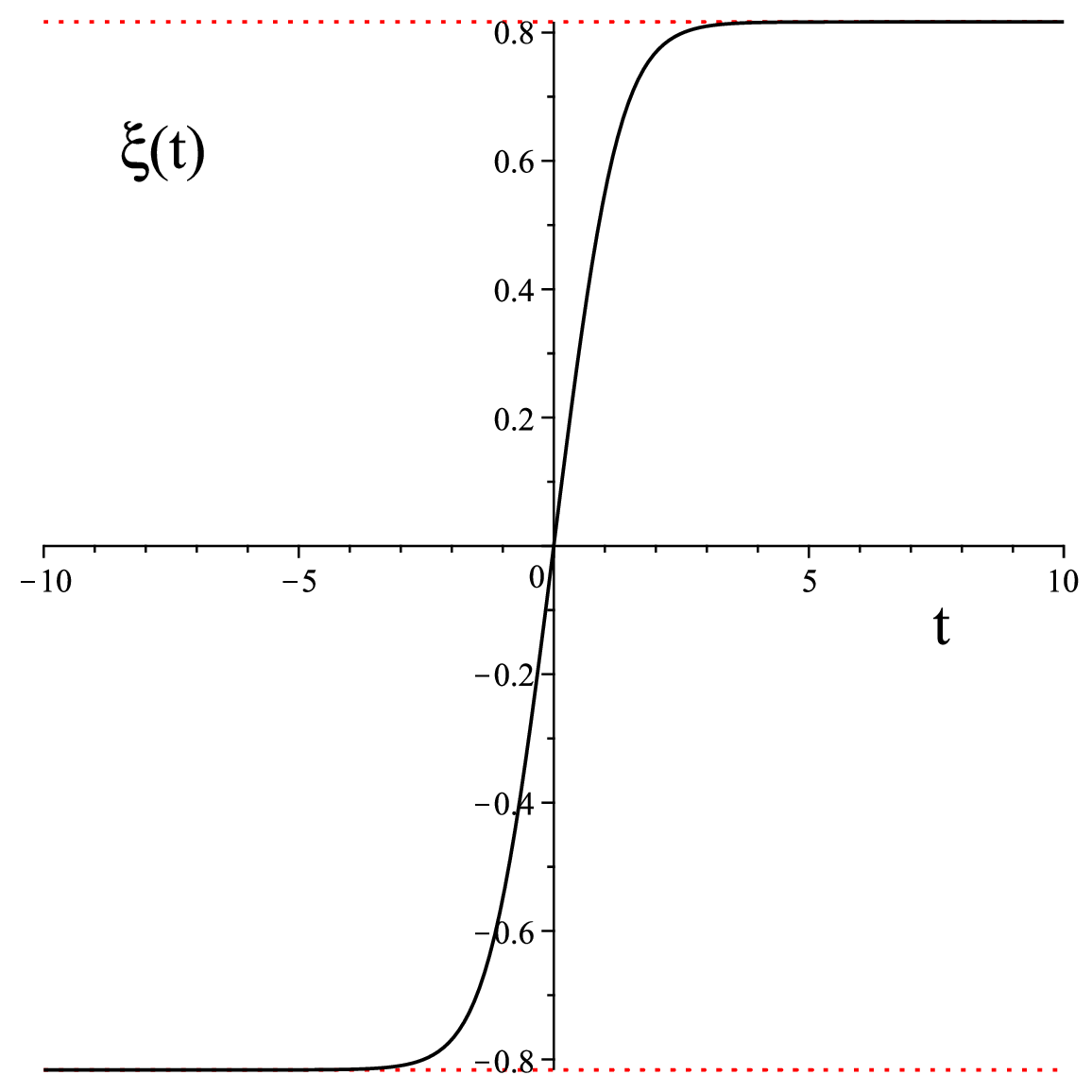}
\caption{
\label{fig:3} 
Plot of of the function $\xi(t)$, Eq. \eqref{xi_def}, including the asymptotes (red 
online, dotted horizontal lines) $\pm \frac{1}{\sqrt{m}}$, for the value $m=3/2$.
}
\end{figure}

By virtue of the $r$ and $t$ equations one finds the ordinary differential equation
\begin{equation}
\frac{dr}{dt}=\frac{C_1}{\cosh(t)\sqrt{C_2\cosh^2(t)+C_1^2}}\,,
\label{(2.28)}
\end{equation}
which can be solved exactly 
\begin{eqnarray}
\label{fin_sol_r}
r(t)=r_* +\frac12 {\rm arctan}\left( {\mathcal R}(t) \right),
\end{eqnarray}
with
\begin{equation}
{\mathcal R}(t)= \frac{(C_1^2-C_2)\cosh^2 t-2C_1^2}{2C_1 \sinh t \sqrt{C_2\cosh^2(t)+C_1^2}},
\label{(2.30)}
\end{equation}
actually depending on the single parameter $m=C_2/C_1^2$ (here assumed positive, for simplicity),
\begin{equation}
{\mathcal R}(t)= \frac{(1-m)\cosh^2 t-2}{2\sinh t \sqrt{m\cosh^2(t)+1}},
\label{(2.31)}
\end{equation}
with
\begin{eqnarray}
{\mathcal R}(t)\big|_{t\to 0}&=&-\frac{\sqrt{m+1}}{2t} + O(t),
\nonumber\\
{\mathcal R}(t)\big|_{t\to \infty}&=&-\frac{(m - 1) }{2\sqrt{m}}+O(e^{-2t}).
\label{(2.32)}
\end{eqnarray}
The solution \eqref{fin_sol_r} can be cast in a more concise form, i.e.,  
\begin{equation}
r(t)=r_* +\frac12 {\rm arctan}\left[\frac12 \left(\xi(t)-\frac{1}{\xi(t)}\right)\right],
\label{(2.33)}
\end{equation}
with 
\begin{equation}
\label{xi_def}
\xi(t)=\frac{ \sinh  t}{ \sqrt{m\cosh^2(t)+1}},
\end{equation}
see Fig. \ref{fig:3}. 

In Eq. \eqref{(2.33)} $r_*$ is an integration constant which can be fixed at $t\to \infty$.
In fact
\begin{equation}
\lim_{t\to \infty}\xi(t)=\frac{1}{\sqrt{m}},
\label{(2.35)}
\end{equation}
and
\begin{equation}
r_\infty =\lim_{t\to \infty} r=r_* +\frac12 {\rm arctan}\left[\frac12 
\left(\frac{1}{\sqrt{m}}-\sqrt{m}\right)\right] ,
\label{(2.36)}
\end{equation}
with ${\rm arctan}\left[\frac12 \left(\frac{1}{\sqrt{m}}-\sqrt{m}\right)\right]$ 
positive when $0<m<1$ (diverging for $m\to 0$, implying $r_\infty>r_*$),
vanishing for $m=1$ (implying $r_\infty=r_*$) and negative for $m>1$ (implying $r_\infty<r_*$). 
Hence
\begin{equation}
m+2\sqrt{m}\tan (2(r_\infty-r_*))-1=0,
\label{(2.37)}
\end{equation}
so that
\begin{equation}
\sqrt{m}=\frac{1}{1+ \sin (2(r_\infty-r_*))}. 
\label{(2.38)}
\end{equation}
Therefore, when $\tau$ (i.e., $t$) spans the full real axis the radial variable 
$r$ is generally bounded, and passes from a minimal to a maximal value.

A final comment concerns the solution \eqref{(2.33)}.
$r(t)-r_*$ changes its sign when $\xi=1$ ($t={\rm arccosh}\sqrt{\frac{2}{1-m}}$) 
and it is singular when $\xi=0$ ($t=0$).

Similarly, also the $\theta$-equation
\begin{eqnarray}
\frac{d\theta}{d\phi}&=& \frac{\pm \sqrt{C_3-\frac{L^2}
{\sin^2\theta}}}{\frac{L}{\sin^2 \theta }},
\label{(2.39)}
\end{eqnarray}
can be integrated exactly. Let us define
\begin{equation}
L=\sqrt{C_3}\sin\theta_* ,
\label{(2.40)}
\end{equation}
and assume, for simplicity, $L,C_3>0$ and $\theta_*\le \theta\le\pi-\theta_* $, 
with $\theta_*\in [0,\pi]$. We then find 
\begin{equation}
\phi(\theta)=\mp {\rm arctan}\left(\frac{\sin\theta_*\cos(\theta)}{\sqrt
{\sin^{2}(\theta)-\sin^{2}\theta_*}}\right),
\label{(2.41)}
\end{equation}
implying that for $\theta=\theta_*,\pi-\theta_*$ one has $\phi=\mp \frac{\pi}{2}$, 
for $\theta=\frac{\pi}{2}$ one has $\phi=0$.
Therefore, when $\phi$ spans the interval $[-\frac{\pi}{2},\frac{\pi}{2}]$ the $\theta$ 
variable is generally bounded, and passes from a minimal 
($\theta_*$) to a maximal value ($\pi-\theta_*$).
To complete the task one has to derive the relations $t=t(\tau)$ 
and $\phi=\phi(\tau)$.
This also follows straightforwardly. For example, upon assuming $t(0)=0$ 
and $\tau(\theta=\frac{\pi}{2})=0$ one has 
\begin{widetext}
\begin{eqnarray}
t(\tau)
&=&\frac{1}{C_1\sqrt{m}}\, 
{\rm arctanh}\left(\sqrt{m}\xi\right)\,,\nonumber\\
\tau(\theta )&=&\mp\frac{1}{\sqrt{C_3}}{\rm arctan}
\left(\frac{\cos\theta}{\sqrt{\sin^2\theta-\sin^2\theta_*}}\right).  
\label{(2.42)}
\end{eqnarray}
\end{widetext}
The fact that in the radial direction the geodesics approach a constant limiting  
value $r_\infty$ with zero speed, $\frac{dr}{d\tau}|_{t\to \infty}\to 0$ (as a consequence 
of Eqs. \eqref{(2.25)}, evaluated in the limit $t\to \infty$) could be expected on the 
basis of general results \cite{Bini:2014esa}. Indeed, the spacetime is getting  
larger and larger in the radial direction as soon as time increases 
\begin{equation}
g_{rr}= \cosh^2 t \bigg|_{t\to \infty}\to \infty ,
\label{(2.43)}
\end{equation}
and this feature forces the geodesics to stop in that direction (i.e., 
to reach the zero-speed limit) \cite{Bini:2014esa}.

Our detailed analysis of the (exact) integrability of geodesics  
in Nariai spacetime cannot be found in the literature, while a discussion
of these aimed at characterizing Singe's world function can be found in Refs. \cite{Wardell,Casals}.

\section{Properties of the Nariai solution in de Sitter-like coordinates}

Let us consider the geodesic equations with respect to the static metric 
in Eqs. \eqref{(1.4)} and \eqref{(1.5)}.
On using the Killing symmetries and denoting by $\bar E$ and $\bar L$ the conserved energy 
and angular momentum (the bar here is needed to avoid confusion with the analogous 
quantities defined in the spherical-like coordinate system) 
\begin{equation}
\frac{dT}{d\tau}= \frac{\bar E}{(1-\rho^2)},
\qquad \frac{d\phi}{d\tau}= \frac{\bar L}{\sin^2\theta}, 
\label{(3.1)}
\end{equation}
the equations are separated and given by
\begin{eqnarray}
\label{geo_gen}
\frac{d\rho}{d\tau}&=&\pm \sqrt{\bar E^2-(1-\rho^2)(1+C_\theta)}\,,\nonumber\\
\left(\frac{d\theta}{d\tau}\right)^2 &=& -\frac{\bar L^2}{\sin^2\theta}+C_\theta.
\end{eqnarray}
For example, in the equatorial case $\theta=\pi/2$, $C_\theta=\bar L^2$ and 
the radial equation reduces to
\begin{eqnarray}
\frac{d\rho}{d\tau}&=&\pm \sqrt{\bar E^2-(1-\rho^2)(1+\bar L^2)}
\nonumber\\
&=& \pm \sqrt{1+\bar L^2}\sqrt{\frac{\bar E^2}{(1+\bar L^2)}-1+\rho^2}.
\label{(3.3)}
\end{eqnarray}
In order to solve the radial equation in Eqs. \eqref{geo_gen} it is convenient to define
\begin{equation}
\alpha^2=\frac{\bar E^2}{1+C_\theta}-1 ,
\label{(3.4)}
\end{equation}
and, assuming $\alpha>0$, rescale both $\tau$ and $\rho$ as
\begin{equation}
\bar \tau=\sqrt{{1+C_\theta}} \; \tau ,
\qquad \bar \rho=\frac{\rho}{\alpha},
\label{(3.5)}
\end{equation}
so that
\begin{equation}
\label{radial_bis}
\frac{d\rho}{d\tau}=\pm \sqrt{1+\bar L^2}\sqrt{\alpha^2+\rho^2},
\end{equation}
becomes
\begin{equation}
\frac{d\bar \rho}{d\bar \tau}=\pm \sqrt{1+ \bar \rho^2}\,,
\label{(3.6)}
\end{equation}
with solution
\begin{equation}
\bar \tau=\pm {\rm arcsinh}(\bar \rho)+C\,,
\label{(3.7)}
\end{equation}
i.e.,
\begin{equation}
\rho(\tau)=\pm \alpha \sinh (\sqrt{1+C_\theta} \tau-C)\,,
\label{(3.8)}
\end{equation}
$C$ being an integration constant which can be set to zero with the choice $\rho(\tau=0)=0$.
When $\alpha=0$, i.e.,
\begin{equation}
\frac{\bar E^2}{1+\bar L^2}=1,
\end{equation}
the radial equation \eqref{radial_bis} simplifies as
\begin{eqnarray}
\frac{d\rho}{d\tau} &=& \pm \sqrt{1+\bar L^2} |\rho|,
\label{(3.3bis)}
\end{eqnarray}
and it can be integrated as 
\begin{equation}
\rho(\tau)= \rho(0)e^{\pm \sqrt{1+\bar L^2}\, \tau }.
\end{equation}
This very simple solution of the equatorial motion will be used later in 
Sec. \ref{test_fluid}, with $\bar L=1$ and $\bar E=\sqrt{2}$.
In addition the $T$ and $\phi$ equations imply
\begin{equation}
\phi(\tau)=\tau+\phi(0)=\tau\,,\qquad \rho(\tau)= \rho(0)e^{\pm \sqrt{2}\, \tau }\,,
\end{equation}
assuming $\phi(0)=0$ and $\rho(0)=\frac12$ (just a simple choice of initial conditions) 
\begin{equation}
\frac{dT}{d\tau}= \frac{2}{1-\frac{e^{\pm 2 \sqrt{2}\, \tau }}4 },
\end{equation}
that is (see Fig. 3)
\begin{equation}
T(\tau)=\mp \frac{1}{2}\ln \left(\frac{4-e^{\pm 2\sqrt{2}\, \tau}}{3} \right)+\sqrt{2}\tau,
\end{equation}
with $T(0)=0$.
Finally, if $\alpha^2<0$, say $\alpha^2=-\beta^2$, Eq. \eqref{radial_bis} becomes
\begin{equation}
\label{radial_bis2}
\frac{d\rho}{d\bar \tau}=\pm \sqrt{\rho^2-\beta^2},
\end{equation}
and the rescaling $\rho=\beta \bar \rho$ implies
\begin{equation}
\label{radial_bis2}
\frac{d\bar\rho}{d\bar \tau}=\pm \sqrt{\bar\rho^2-1},
\end{equation}
that is
\begin{equation}
\bar\rho=\pm \sin \bar \tau.
\end{equation}

Let us now revert to the $\theta$ equation, which 
similarly, can be integrated as follows. Let us re-write it as
\begin{equation}
\frac{d\theta}{d\tau}=\pm\frac{1}{\sin \theta}\sqrt{C_\theta  \sin^2\theta -\bar L^2}.
\label{(3.9)}
\end{equation}
Upon passing to $z=\cos\theta$ we find
\begin{equation}
\frac{dz}{d\tau}= \mp \sqrt{C_\theta}\sqrt{\frac{C_\theta-\bar L^2}{C_\theta}-z ^2 },
\label{(3.10)}
\end{equation}
that is
\begin{eqnarray}
\mp \sqrt{C_\theta} \tau &=& \int \frac{dz}{\sqrt{A^2-z ^2 }}
\nonumber\\
&=& {\rm arctan}\left( \frac{z}{\sqrt{A^2-z ^2}}\right)+C,
\label{(3.11)}
\end{eqnarray}
where 
\begin{equation}
A^2=\frac{C_\theta-\bar L^2}{C_\theta},
\label{(3.12)}
\end{equation}
and $C$ is an integration constant, that we set to zero by choosing  
$z(0)=0$, i.e., $\theta(0)=\frac{\pi}{2}$. Finally
\begin{equation}
\theta(\tau)={\rm arccos}\left( \sqrt{\frac{C_\theta-\bar L^2}{C_\theta}}
\sin(C_\theta \tau) \right).
\label{(3.13)}
\end{equation}
The solutions for $T(\tau)$ and $\phi(\tau)$ then follow easily.

\begin{figure}
\includegraphics[scale=0.3]{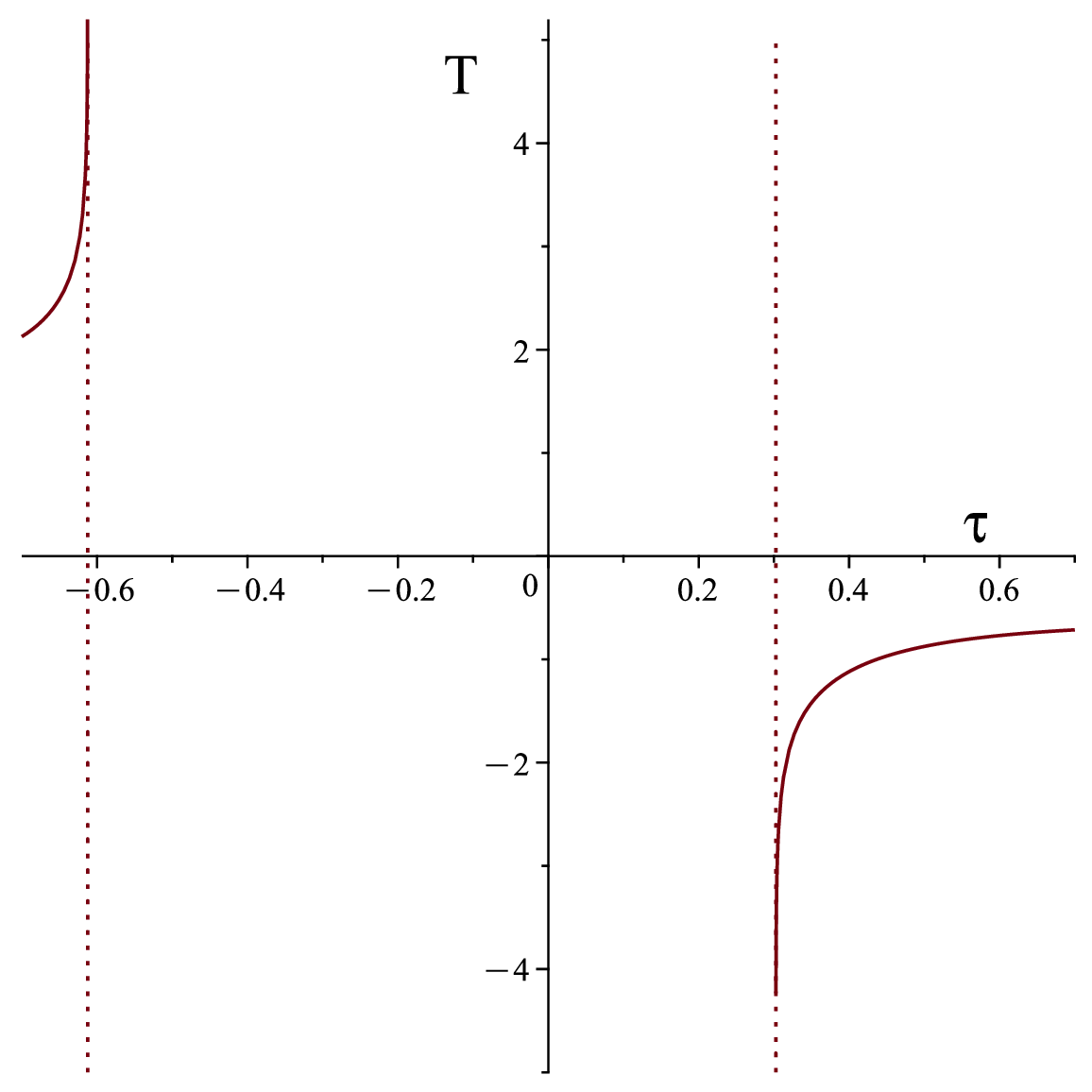}
\includegraphics[scale=0.3]{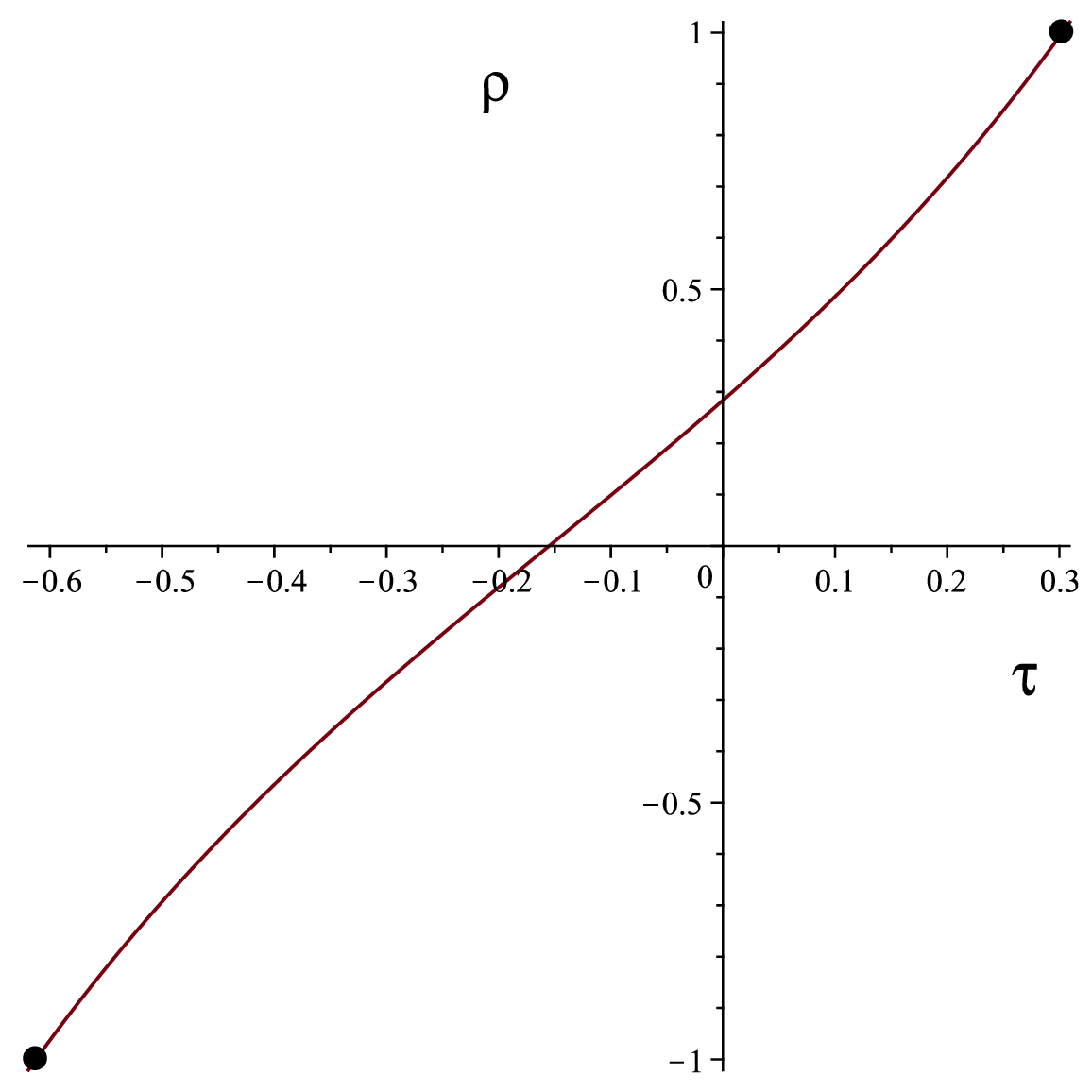}
\caption{
\label{fig:rhoT} 
Plot of $T(\tau)$ (upper plot) and $\rho(\tau)$ (lower plot) obtained by using 
Eq. \eqref{(1.6)} and the geodesics of Fig. \eqref{fig:1}.
In the plot $T(\tau)$ the two vertical asymptotes occur at $\tau=-0.6124226908$ 
($\sin r/\tanh t=1$) and $\tau=0.3025413974$ ($\sin r/\tanh t=-1$). Black dots 
correspond to $\rho=\pm 1$, i.e., the maximum and minimum allowed values for $\rho$. 
}
\end{figure}

It is worth noting that equatorial ($\theta=\frac{\pi}{2}$) circular geodesics 
($\rho=\rho_0$, fixed) exist for $C_\theta=\bar L^2$. In this case we have
\begin{eqnarray}
T&=&\frac{\bar E}{(1-\rho_0^2)}\tau\equiv \Gamma \tau , 
\nonumber\\
\phi&=&\bar L \tau ,
\label{(3.14)}
\end{eqnarray}
with the constraint
\begin{equation}
1-\rho_0^2=\frac{\bar E^2}{1+\bar L^2},\qquad \to \qquad \rho_0
=\sqrt{\frac{1+\bar L^2-\bar E^2}{1+\bar L^2}}.
\label{(3.15)}
\end{equation}
Unfortunately, the causality condition according to which 
these orbits should be timelike (a case of special interest here) imposes
$E^{2}=L^{2}+1$, implying that they exist only at $\rho=0$.

We conclude this section by pointing out that the three-dimensional 
Nariai metric induced on the $T=$constant slices
\begin{equation}
{}^{(3)}ds^2=\frac{d\rho^2}{(1-\rho^2)}+d\theta^2+\sin^2\theta d\phi^2
\end{equation}
is conformally flat. Indeed, the associated Cotton-York tensor vanishes identically
\begin{equation}
{}^{(3)}R_{a[b;c]}=0\,,\qquad a,b,c=1,2,3.
\end{equation}
In fact,  changing the temporal coordinate (only) as
\begin{equation}
{\mathcal T}=T-\frac12 \ln(1-\rho^2)
\label{(3.28)}
\end{equation}
in the original Nariai metric, Eqs. \eqref{(1.4)} and \eqref{(1.5)}, one obtains
\begin{eqnarray}
ds^2&=&-(1-\rho^2)d{\mathcal T}^2+2\rho d{\mathcal T} d\rho +d\rho^2
+d\theta^2+\sin^2\theta d\phi^2
\nonumber\\
&=& -d{\mathcal T}^2+ (d\rho+\rho d{\mathcal T})^2+d\theta^2+\sin^2\theta d\phi^2.
\end{eqnarray}
The induced metric on the slices ${\mathcal T}$=constant is given by
\begin{equation}
ds^2|_{{\mathcal T}={\rm const}}=d\rho^2+d\theta^2+\sin^2\theta d\phi^2,
\end{equation}
and it is conformally flat \lq\lq at sight," i.e.,
\begin{eqnarray}
d\rho^2+d\theta^2+\sin^2\theta d\phi^2&=&\frac{1}{R^2}[dR^2
\nonumber\\
&+& R^2 d\theta^2+\sin^2\theta d\phi^2],
\qquad
\end{eqnarray}
with
\begin{equation}
\rho=\ln R.
\end{equation}
The relation \eqref{(3.28)} can be seen as defining the so-called 
Painlev\'e-Gullstrand observers of this spacetime
\begin{equation}
-d{\mathcal T}=-dT-\frac{\rho}{(1-\rho^2)}d\rho,
\end{equation}
with dual
\begin{equation}
\partial_{\mathcal T}=\frac{1}{(1-\rho^2)}\partial_T -\rho \partial_\rho. 
\end{equation}
The four-velocity of these observers is exactly $\partial_{\mathcal T}$ 
($\partial_{\mathcal T}\cdot \partial_{\mathcal T}=-1$) and corresponds 
to geodesic (radially inward) motion.

\section{Accelerated motions}

We will discuss below two examples of accelerated motion: the one of 
observers at rest with respect to the coordinates (static observers) 
and the one of observers in radial motion with respect to the static 
family of observers. For both cases we will investigate the geometric 
characterization of the four-velocity in terms of vorticity and expansion, 
besides the mentioned acceleration.

More specifically, we will explore now special types of motion with respect to 
both coordinate systems $(t,r,\theta,\phi)$ (here\-after spherical-like 
coordi\-nates) and $(T,\rho,\theta,\phi)$ (here\-after dS-like coordinates), 
i.e., those of a particle at rest and in radial motion.

\subsection{At rest with respect to spherical-like coordinates in Nariai spacetime} 

In order to better understand the features of the gravitational field in Nariai spacetime, 
let us consider an observer at rest with respect to the coordinates, 
i.e., with four-velocity
\begin{equation}
u=\partial_t\,, \qquad u^\flat =-dt.
\label{(4.1)}
\end{equation}
$u$ is geodesic and irrotational (in the sense that, for $u$, acceleration 
and vorticity both vanish identically).
However, the associated congruence is expanding, with 
\begin{equation}
u^{\alpha}{}_{;\beta}=\sigma \, \delta^\alpha_r \delta_\beta^r ,
\label{(4.2)}
\end{equation}
(automatically orthogonal to $u$) and with shear 
\begin{equation}
\sigma=u^\alpha{}_{;\alpha}=\tanh t,
\label{(4.3)}
\end{equation}
i.e., changing sign (vanishing) at $t=0$, positive for $t>0$ and negative for $t<0$. 
The situation of constant shear $\sigma=\pm 1$ is approached only asymptotically.

\subsection{Radial motion with respect to spherical-like coordinates in Nariai spacetime} 

Let $U$ be an observer moving radially in Nariai spacetime, with four-velocity
\begin{equation}
U=\gamma (u+v e_1),\quad \gamma =(1-v^2)^{-1/2} ,
\label{(4.4)}
\end{equation}
where $v$ is a constant (with its sign).
We find for $U$ a nonvanishing acceleration
\begin{equation}
a(U)= \gamma^2 v \tanh(t) \left( v \partial_t+ e_1 \right) ,
\label{(4.5)}
\end{equation}
By introducing the boosted (spatial) radial vector
\begin{equation}
E_{1}(U)=\gamma \left( v \partial_t+ e_1 \right),
\label{(4.6)}
\end{equation}
of unit norm and orthogonal to $U$, we find
\begin{equation}
a(U)= \kappa(t)  E_{1}(U) ,
\label{(4.7)}
\end{equation}
with 
\begin{equation}
\kappa(t)= \gamma  v\tanh t ,
\label{(4.8)}
\end{equation}
the signed magnitude of the four-acceleration.
The maximal value of the acceleration as a function of $t$ 
is obtained as soon as $t\to \infty$,
\begin{equation}
\lim_{t\to \pm \infty}\kappa(t)=\pm \gamma  v .
\label{(4.9)}
\end{equation}
The minimal value is instead the geodesic one, reached at $t=0$.

\subsection{At rest with respect to de Sitter-like coordinates in Nariai spacetime} 

Let us recall the relations
\begin{equation}
\rho=\cosh t \cos r ,\qquad T={\rm arctanh}\left(\frac{\cosh t \sin r}{\sinh t}\right) ,
\label{(4.10)}
\end{equation}
and let us bear in mind the dS-like form \eqref{(1.4)} and \eqref{(1.5)} of the metric.
Clearly, an observer at $r=r_0$ fixed moves radially in the $\rho$-direction, 
with the radial coordinate increasing continuously
\begin{equation}
\rho=\cosh t \cos r_0,\qquad T={\rm arctanh}\left(\frac{\cosh t \sin r_0}{\sinh t}\right) ,
\label{(4.11)}
\end{equation}
that is
\begin{equation}
t={\rm arccosh}\left(\frac{\rho}{\cos r_0}\right)\,,\qquad \tanh T= 
\frac{\rho \tan r_0}{\sqrt{\frac{\rho^2}{\cos^2 r_0}-1}} .
\label{(4.12)}
\end{equation}

An observer at rest with respect to these coordinates has four-velocity
\begin{equation}
v=\frac{1}{\sqrt{1-\rho^2}}\partial_T , 
\qquad v^\flat =-\sqrt{1-\rho^2} dT ,
\label{(4.13)}
\end{equation}
and radial acceleration
\begin{equation}
a(v)=-\rho \partial_\rho.
\label{(4.14)}
\end{equation}
Moreover
\begin{equation}
v^{\alpha}{}_{;\beta}=-\rho \sqrt{1-\rho^2}\, \delta^\alpha_T \delta_\beta^\rho ,
\label{(4.15)}
\end{equation}
(automatically orthogonal to $v$) so that the congruence associated with $v$ turns 
out to be vorticity-free and expansion-free.

\subsection{Radial motion with respect to de Sitter-like coordinates in Nariai spacetime} 

Let us introduce the two unit vectors
\begin{equation}
e_0=\frac{1}{\sqrt{1-\rho^2}}\partial_T,\qquad e_1=\sqrt{1-\rho^2}\partial_\rho ,
\label{(4.16)}
\end{equation}
and the corresponding two vectors boosted along the radial direction
\begin{equation}
V=\gamma (e_0 +v e_1),\qquad  E_1(V)=\gamma (ve_0+e_1) .
\label{(4.17)}
\end{equation}
$V$ represents the four-velocity of a test particle moving radially.
It is accelerated radially inwards with acceleration
\begin{equation}
a(V)=-\frac{\gamma \rho}{\sqrt{1-\rho^2}}E_1(V).
\label{(4.18)}
\end{equation}

\section{Massless scalar perturbations of Nariai spacetime in absence of sources}

Vacuum scalar perturbations of Nariai spacetime are summarized by the Klein-Gordon 
equation which, with our choice of signature ($-+++$), takes the form
\begin{equation}
(\Box -m^2 -\xi R) \psi=0,
\label{(5.1)}
\end{equation}
where $\Box = \nabla^\mu \nabla_\mu$ and $\xi=1/6$ for a conformally invariant wave equation.
In this case, since the Ricci scalar is constant as well, we can study the equation
\begin{equation}
\label{KG}
(\Box  -\mu^2) \psi=0,
\end{equation}
where
\begin{equation}
\mu^2=m^2+\xi R ,
\label{(5.3)}
\end{equation}
is constant, representing an \lq\lq effective mass" term including curvature contributions.
For example, in the massless case ($m=0$) and conformally invariant equation 
($\xi=1/6$) we have $\mu^2=\frac23$. 
Since separability of a differential equation is a matter of coordinate 
it is worth studying this equation in both coordinate systems 
$(t,r,\theta,\phi)$ (with a $t$-dependent metric) and $(T,\rho,\theta,\phi)$ 
(with a $\rho$-dependent metric) introduced above. Moreover, since only the 
$(t,r)$, $(T,\rho)$ sections are modified by the coordinate transformation 
\eqref{(1.6)}, the angular coordinates in both cases separate in terms of 
spherical harmonics due to the underlying spherical symmetry of the $(\theta,\phi)$ sector.

A final remark concerns the Fourier transform of the field variable: in one set 
of coordinates, $(t,r,\theta,\phi)$ it is natural to transform the radial variable, 
whereas in the other set, $(T,\rho,\theta,\phi)$, it is natural to transform the 
temporal variable. This makes the discussion of both cases different and, 
at the same time, interesting as well. At a deeper level, in a general curved
spacetime, the homogeneity required for the existence of a momentum-space 
representation is lacking \cite{BunchParker}, and one has to resort to the use
of Fourier-Maslov integral operators \cite{Treves}. However, when some symmetries
exist, one can still use the tool of Fourier transform. For example, the scalar
wave equation in Schwarzschild spacetime can be solved by taking the Fourier
transform with respect to the time variable \cite{Persides}, because in
$(t,r,\theta,\phi)$ coordinates the metric is independent of $t$.

\subsection{Scalar waves in spherical-like coordinates}

By virtue of the previous remarks, since the Nariai metric \eqref{(1.1)}
is independent of the $r$ variable, we can introduce a Fourier transform 
with respect to this variable, looking therefore 
for separable solutions of Eq. \eqref{KG} in the factorized form
(hereafter, $l=0,1,...,\infty$ and $m=-l,-l+1,...,l$)
\begin{equation}
\label{psi_sol}
\psi(t,r,\theta,\phi)=\sum_{lm}\int \frac{d\omega}{2\pi}e^{+i\omega r}
T_{lm\omega}(t) Y_{lm}(\theta,\phi),
\end{equation}
so that
\begin{eqnarray}
\Box \psi&=&\sum_{lm}Y_{lm}\int \frac{d\omega}{2\pi}e^{+i\omega r} \left[-T_{lm\omega}''
-\tanh t T_{lm\omega}'\right.
\nonumber\\
&&\left.-\frac{\omega^2}{\cosh^2t}T_{lm\omega}-L T_{lm\omega}  \right] ,
\label{(5.5)}
\end{eqnarray}
where $L=l(l+1)$. Equation \eqref{KG} then implies
\begin{equation}
\label{eqT_fin}
T_{lm\omega}''+\tanh t T_{lm\omega}' +\left(\frac{\omega^2}{\cosh^2t} 
+L  +\mu^2\right) T_{lm\omega}=0 .
\end{equation}
Actually, here $T_{lm\omega}$ does not depend on $m$: we thus simplify the notation denoting 
it simply by $T_{l\omega}(t)$; moreover, we introduce the notation
\begin{equation}
L  +\mu^2=L+ m^2+\xi R={\mathcal L} ,
\label{(5.7)}
\end{equation}
since it is only this combination which occurs in Eq. \eqref{eqT_fin}. Hence, Eq. \eqref{eqT_fin} becomes 
\begin{equation}
\label{eq_T}
T_{l\omega}''+\tanh t T_{l\omega}' +\left(\frac{\omega^2}{\cosh^2t} 
+{\mathcal L}\right) T_{l\omega}=0 .
\end{equation}
At this stage, we express $T_{l\omega}(t)$ in the form
\begin{equation}
T_{l\omega}(t)=\frac{1}{\sqrt{\cosh t}} S_{l\omega}(t),
\label{(5.9)}
\end{equation}
which brings Eq. \eqref{eq_T} to its normal form
\begin{equation}
\label{eq_S}
S_{l\omega}''- \left(\frac{\frac{1-4\omega^2}{4}}{\cosh^2t} 
+\frac{1-4{\mathcal L}}{4}\right) S_{l\omega}=0.
\end{equation}
Inspection of Eq. \eqref{eq_S} shows that it is convenient to introduce the linear function
\begin{equation}
a(x)=\frac{1-4x}{4},
\label{(5.11)}
\end{equation}
so that one can write
\begin{equation}
\label{eq_Sbis}
S_{l\omega}''- \left(\frac{a(\omega^2)}{\cosh^2t} 
+a({\mathcal L})\right) S_{l\omega}=0.
\end{equation}

It is convenient to introduce also the notation
\begin{equation}
\nu^2=\frac{1-4{\mathcal L}}{4}=a({\mathcal L})
\,,\qquad \omega=\frac12+\tilde l.
\label{(5.13)}
\end{equation}
In this way
\begin{equation}
a(\omega^2)=-\tilde l (\tilde l+1)=-\tilde L \,,
\label{(5.14)}
\end{equation}
where we have set
\begin{equation}
\tilde L=\tilde l(\tilde l+1),
\label{(5.16)}
\end{equation}
in analogy with  the above defined quantity $L=l(l+1)$.
Thus, Eq. \eqref{eq_S} becomes
\begin{equation}
\label{eq_S2}
S_{l\omega}''(t)+ \left(\frac{\tilde L}{\cosh^2t}-\nu^2\right) S_{l\omega}(t)=0\,.
\end{equation}

Last, upon passing to the new variable
\begin{equation}
u=\tanh t\,, \qquad u\in (-1,1)\,,
\label{(5.17)}
\end{equation}
and denoting by a dot the derivative with respect to $u$, the above equation \eqref{eq_S2} becomes
\begin{eqnarray}
&&(1-u^2)^2\ddot S_{\tilde l \nu}(u) -2 u (1-u^2)\dot S_{\tilde l \nu}(u) 
\nonumber\\ 
&&\qquad +[\tilde L (1-u^2)-\nu^2]S_{\tilde l \nu}(u)=0,
\label{(5.18)}
\end{eqnarray}
and its general solution is expressed in terms of associated Legendre polynomials
\begin{equation}
S_{\tilde l \nu}(u)=C_1 P_{\tilde l, \nu}(u)+C_2 Q_{\tilde l, \nu}(u)\,,
\label{(5.19)}
\end{equation}
here \lq\lq generalized" since the indices $(\tilde l, \nu)$ are not necessarily 
integers and expressed in terms of hypergeometric functions, i.e., 
\begin{eqnarray}
\label{PQ_defs}
P_{\tilde l, \nu}(u)&=& \frac{1}{\Gamma(1-\nu)}\left(\frac{1-u}{1+u}\right)^{\nu/2}\times 
\nonumber\\
&&{}_2F_1\left(-\tilde l, 1+\tilde l, 1+\nu, \frac{1-u}{2} \right),
\nonumber\\
Q_{\tilde l, \nu}(u)&=& \frac{\sqrt{\pi}}{2^{\tilde l +1}}\frac{\Gamma(\tilde l+\nu+1)}
{\Gamma(\tilde l+\frac32)}\frac{(1-u^2)^{\nu/2}}{u^{\tilde l +\nu +1}}\times 
\nonumber\\
&&{}_2F_1\left(\frac12 (\tilde l+\nu+1), \frac12 (\tilde l+\nu+2), 
\tilde l+\frac32, \frac{1}{u^2} \right),
\nonumber\\
\end{eqnarray}
where from Eqs. \eqref{(5.7)} (with $m=0=\xi$) and \eqref{(5.13)}\,,
\begin{equation}
\tilde l=\omega-\frac12\,,\qquad \nu=\sqrt{\frac14 -l(l+1)}\,,
\end{equation}
i.e., $\tilde l$ is a function of the continuous parameter $\omega$ and 
$\nu$ is a complex function of $l=0,1,2,\ldots$ (see Fig. 4).
Consequently, a resummed, exact expression for the field, \eqref{psi_sol} can only be formal.

\begin{figure}
\includegraphics[scale=0.35]{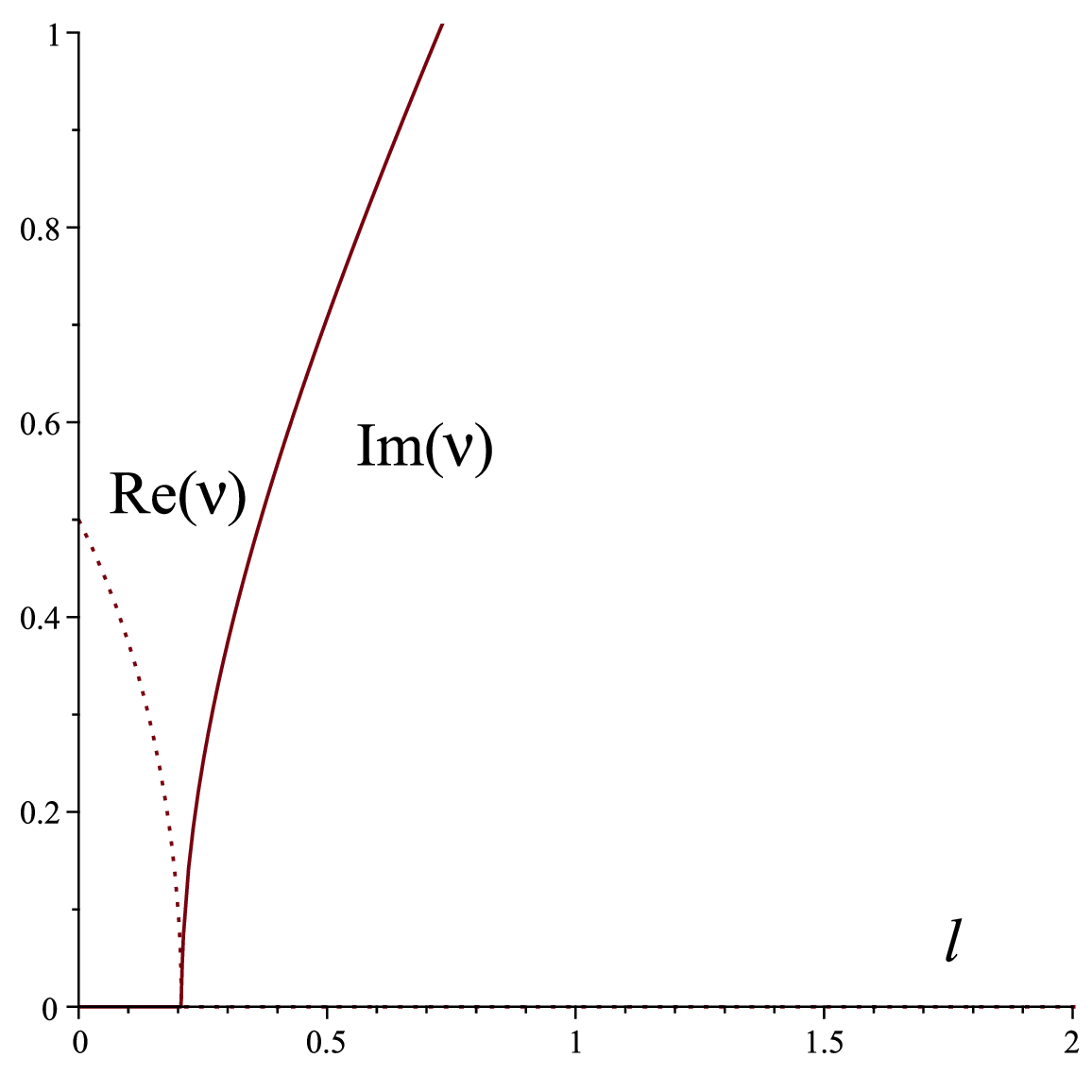}
\caption{\label{re_im_part_nu} The real and imaginary parts of $\nu$ as a function of $l$, 
assumed as a continuous parameter for graphical reasons. The cexplicit values for 
$l=0\ldots 10$ are $\nu$=[0.5, 1.3229\, i, 2.3979\, i, 3.4278\, i, 4.4441\, i, 5.4544\, 
i, 6.4614\, i, 7.4666\, i, 8.4705\, i, 9.4736\, i, 10.4762\, i], respectively.}
\end{figure}
The constants should be properly chosen to have $S_{\tilde l \nu}(u)$, Eq.
\eqref{(5.19)}, a real function.
Let us notice, incidentally, that the generalized Legendre polynomials 
are also called Legendre functions in the literature.

As we will also see below, these pair of functions $P_{\tilde l, \nu}(u)$ and 
$Q_{\tilde l, \nu}(u)$ are two independent solutions of Eq. \eqref{eq_T}. 
Similarly, nontrivial combinations of these solutions can be taken as independent solutions as well. 
Usually, the independence 
of two solutions is checked by looking at their dependence upon the parameters 
occurring in Eq. \eqref{eq_T}. For example, the latter equation depends on 
$\tilde l$ via the combination $\tilde l (\tilde l+1)$, which is unaffected 
by the map $\tilde l \to -\tilde l -1$. Thus, two independent solutions 
are $P_{\tilde l, \nu}(u)$ and $P_{-\tilde l-1, \nu}(u)$. Similarly, it 
depends also on $\nu$ via the combination $\nu^2$, which is unaffected by 
the map $\nu \to -\nu$. Hence two independent solutions are 
$P_{\tilde l, \nu}(u)$ and $P_{\tilde l, -\nu}(u)$, etc.
 
Last, the expansion \eqref{psi_sol} becomes
\begin{equation}
\psi(t,r,\theta,\phi)=\sum_{lm} \frac{Y_{lm}(\theta,\phi)}{\sqrt{\cosh t}}K_{l\omega}(t,r),
\label{(5.21)}
\end{equation}
where
\begin{equation}
K_{l\omega}(t,r)=\int \frac{d\omega}{2\pi}e^{i\omega r} S_{\omega-\frac12, 
\frac{\sqrt{1-4{\mathcal L}}}{2}}(\tanh t).
\label{(5.22)}
\end{equation}
One can limit considerations to the $\omega$ space and perform the Fourier 
transform (with respect to $r$) of the scalar field
\begin{equation}
\hat \psi(t,\omega,\theta,\phi)=\sum_{l} \frac{Y_{l0}(\theta,\phi)}
{\sqrt{\cosh t}}S_{\omega-\frac12, 
\frac{\sqrt{1-4{\mathcal L}}}{2}}(\tanh t),
\label{(5.23)}
\end{equation}
which does not depend on $m$, in some special situation, e.g. $\theta=\frac{\pi}{2}$, $\phi=0$,
\begin{eqnarray}
\hat \psi(t,\omega,\frac{\pi}{2},0)&=&\sum_{l} 
\frac{Y_{l0}(\frac{\pi}{2},0)}{\sqrt{\cosh t}}S_{\omega-\frac12, 
\frac{\sqrt{1-4{\mathcal L}}}{2}}(\tanh t)
\nonumber\\
&=& \sum_{l} \frac{ \sqrt{2 l + 1} S_{\omega-\frac12, 
\frac{\sqrt{1-4{\mathcal L}}}{2}}(\tanh t)}{ 2\Gamma(\frac{1-l}{2} )\Gamma(1 
+ \frac{l}{2}) \sqrt{\cosh t}}.\qquad
\label{(5.24)}
\end{eqnarray}
However, even in this simplified case, any further analysis can only be performed numerically.
 
\subsection{Scalar waves in de Sitter-like coordinates}

Because of the Killing properties of the temporal variable $t$, 
let us look for separable solutions in the factorized form
\begin{equation}
\label{psi_sol_Trho}
\psi(T,\rho,\theta,\phi)=\sum_{lm}\int \frac{d\omega}{2\pi}e^{-i\omega T}
R_{lm\omega}(\rho) Y_{lm}(\theta,\phi).
\end{equation}
The Laplacian of $\psi$ yields
\begin{eqnarray}
\Box \psi&=&\sum_{lm}Y_{lm}\int \frac{d\omega}{2\pi}e^{-i\omega T} \left[ 
(1-\rho^2)R_{lm\omega}''-2\rho R_{lm\omega}'\right.
\nonumber\\
&+& \left. \left(\frac{\omega^2}{(1-\rho^2)}-L\right)R_{lm\omega}\right].
\label{(5.26)}
\end{eqnarray}
Equation \eqref{KG} then implies
\begin{eqnarray}
&& R_{lm\omega}''-\frac{2\rho}{(1-\rho^2)} R_{lm\omega}' 
\nonumber\\
&& +\frac{1}{(1-\rho^2)}\left(\frac{\omega^2}{(1-\rho^2)}-{\mathcal L} \right)R_{lm\omega}=0,
\label{(5.27)}
\end{eqnarray}
where the relation ${\mathcal L}$ vs $L$ is given in Eq. \eqref{(5.7)}.

with solution expressed again in terms of associated Legendre polynomials with complex order,
\begin{equation}
R_{lm\omega}(\rho)= C_1 P_{\widehat l, i\omega}(\rho)+C_2 Q_{\widehat l, i\omega}(\rho),
\label{(5.28)}
\end{equation}
where $C_1$ and $C_2$ should be chosen properly,  
\begin{equation}
\widehat l=\frac{(1-4{\mathcal L})^{1/2}}{2}-\frac{1}{2}=\nu -\frac12\,,
\label{(5.29)}
\end{equation}
and the Legendre functions $P_{l,m}(\rho)$, $Q_{l,m}(\rho)$ have been defined in Eq. \eqref{PQ_defs}.  
Note that $\hat l$ is a function of $l$ and hence on the left-hand-side 
of Eq. \eqref{(5.28)} we have left $R_{lm\omega}$.

As in the previous case, one can look at $\psi$ in the $\omega$-space ($\hat \psi$ 
denoting the Fourier transform of $\psi$ with respect to the variable $T$), and 
focusing on special values of $\theta$ and $\phi$, e.g., 
$\theta=\frac{\pi}{2}$, $\phi=0$. In this case
\begin{equation}
\label{psi_sol_Trho_omega_space}
\hat \psi(\omega,\rho,\frac{\pi}{2},0)=\sum_{lm} 
R_{lm\omega}(\rho) Y_{lm}(\frac{\pi}{2},0)\,.
\end{equation}
Also in this case the analysis (e.g., 
the evaluation of gauge invariant quantities, etc.) can be performed only numerically.

\section{Waves sourced by a particle carrying a scalar charge: 
an example of analytic self-force computation}

Studies of scalar or gravitational Self Force (SF) in the Nariai spacetime were 
initiated by Wardell and collaborators \cite{Wardell,Casals}, 
aimed at taking some advantage from this simplified situation with respect to the 
corresponding, much more interesting for what concerns applications, black hole case.

The most common situation implies a scalar charge located along a timelike worldline 
(possibly geodesic) as a source of the perturbation (cf. Ref. \cite{Casals}),
\begin{eqnarray}
\label{scalar_field_eq_source}
(\Box  -\mu^2) \psi&=&-4\pi \rho_{s},
\nonumber\\
\rho_{s}&=&q_s \int \frac{ d\tau}{\sqrt{-g}}\delta^{(4)}(x-z(\tau)),
\end{eqnarray}
with $z^\alpha(\tau)$ the parametric equations of the source orbit,  with  $\tau$ the proper time.

In order to proceed further with the analysis of Eq. 
\eqref{psi_sol_Trho_omega_space}, we need to specify a coordinate choice and the 
orbit of the perturbing particle.
In the $(t,r)$ coordinates it is natural to study the simple case of a particle 
at rest, since it follows a geodesic with four-velocity $u=\partial_t$, 
but the wave equation is much involved.
In the $(T,\rho)$ coordinates a particle at rest, with four-velocity $U=\partial_T$, 
is (radially) accelerated, i.e., its motion is not geodesic. In other words, to keep 
the particle at rest one needs a freely specifiable support, e.g., a rocket. The advantage, 
however, is that the wave equation is not so involved as in the previous case.

\subsection{Particle at rest in the $(t,r)$ coordinates}

Since in this case a particle at rest follows geodesic motion, with 
four-velocity $U=\partial_t$, we will study the scalar perturbations induced on 
Nariai spacetime by such a particle with world line parametrized as
\begin{equation}
t(\tau)= \tau , \qquad r(\tau)=r_0 , \qquad \theta_0=\frac{\pi}{2} , \qquad \phi_0=0,
\label{(6.2)}
\end{equation}
placed on the equatorial plane $\frac{\pi}{2}$, without any loss of generality.
This implies that
\begin{equation}
\rho_{s}=\frac{q_s}{\cosh t}  \delta(r-r_0)\delta(\theta-\frac{\pi}{2})\delta(\phi),
\label{(6.3)}
\end{equation}
since
\begin{equation}
\sqrt{-g}\big|_{\rm orbit}= \cosh(\tau).
\label{(6.4)}
\end{equation}
Let us consider in detail Eq. \eqref{scalar_field_eq_source}, written in the form
\begin{equation}
\left(\Box  -\mu^2\right)  \psi=-4 \pi  \frac{q_s}{\cosh t}  \delta(r-r_0)
\delta(\theta-\frac{\pi}{2})\delta(\phi).
\label{(6.5)}
\end{equation}
As in the source-free case, the angular part of this equation decouples when one 
decomposes the field $\psi$ in spherical harmonics, leading to a Schr\"odinger-like 
radial equation, sourced by a Dirac delta centered at a certain 
value $r_0$ of the radial variable.
Let us look for solutions of the type
\begin{equation}
\psi(t,r,\theta,\phi)=\sum_{lm} \int \frac{d\omega}{2\pi}e^{i\omega (r-r_0)} 
Y_{lm}(\theta,\phi) T_{lm\omega}(t),
\label{(6.6)}
\end{equation}
i.e., adding the constant $e^{-i\omega r_0}$ with respect to the case of absence of sources, 
also recalling that
\begin{equation}
\delta(\theta-\frac{\pi}{2})\delta(\phi)=\sum_{lm}Y_{lm}(\theta,\phi)Y_{lm}^*(\frac{\pi}{2},0).
\label{(6.7)}
\end{equation}
We find
\begin{widetext}
\begin{eqnarray}
&&\Box \psi=\sum_{lm}\int \frac{d\omega}{2\pi}e^{i\omega (r-r_0)}\left\{-[\tanh t  
T_{lm\omega}'(t)+T_{lm\omega}''(t)+ L T_{lm\omega}(t)]-\omega^2
\frac{T_{lm\omega}(t)}{\cosh^2(t)}\right\}Y_{lm}(\theta,\phi),
\nonumber\\
&& -\mu^2 \psi =-\mu^2 \sum_{lm} \int \frac{d\omega}{2\pi}e^{i\omega (r-r_0)} 
T_{lm\omega}(t)   Y_{lm}(\theta,\phi),
\nonumber\\
&& -4 \pi \frac{q_s}{\cosh t}  \delta(r-r_0)\sum_{lm}Y_{lm}(\theta,\phi)
Y_{lm}^*(\frac{\pi}{2},0)
=-\sum_{lm}\frac{S_{lm}}{\cosh t}\int \frac{d\omega}{2\pi}
e^{i\omega (r-r_0)}Y_{lm}(\theta,\phi),
\label{(6.8)}
\end{eqnarray}
where
\begin{equation}
S_{lm}=4\pi q_s Y_{lm}^*(\frac{\pi}{2},0),
\label{(6.9)}
\end{equation}
is a constant. Equation \eqref{(6.5)} then becomes
\begin{equation}
\label{eq_fund}
T_{lm\omega}''(t)+\tanh t  T_{lm\omega}'(t)+\left(\frac{\omega^2}{\cosh^2(t)}
+{\mathcal L}\right)T_{lm\omega}(t)=\frac{S_{lm}}{\cosh t}.
\end{equation}
The general solution of this equation follows by using the Green function method. 
In fact, the two independent solutions of the homogeneous equation are given by
\begin{eqnarray}
T_{l\omega}^{(1)}(t)&=&\frac{P_{\tilde l, \nu}(\tanh (t))}{\sqrt{\cosh t}}\,,
\nonumber\\
T_{l\omega}^{(2)}(t)&=&\frac{Q_{\tilde l, \nu}(\tanh (t))}{\sqrt{\cosh t}}\,,
\label{(6.11)}
\end{eqnarray}
i.e., do not depend on $m$ but only on $l$ and $\omega$.
Their constant Wronskian is
\begin{equation}
W_{l\omega}=\cosh t [T_{l\omega}^{(1)}{}'(t)T_{l\omega}^{(2)}(t)-T_{l\omega}^{(1)}(t)T_{l\omega}^{(2)}{}'(t)],
\label{(6.12)}
\end{equation}
where we recall that $\tilde l$ and $\nu$ are related to $l$ and $\omega$ and given in Eq. 
\eqref{(5.13)}. 
The general solution (depending also on $m$ because of the source term $S_{lm}$) reads then
\begin{eqnarray}
T_{lm\omega}(t)&=&\frac{S_{lm}}{W_{l\omega} \sqrt{\cosh (t)}}\left[T_{l\omega}^{(1)}(t) \int^t dt'  
T_{l\omega}^{(2)}(t') -T_{l\omega}^{(2)}(t) \int^t dt'  T_{l\omega}^{(1)}(t') \right]
+C_1 T_{l\omega}^{(1)}(t)+C_2 T_{l\omega}^{(2)}(t),
\label{(6.13)}
\end{eqnarray}
with $C_{1,2}$ two arbitrary constants.

One can evaluate $\psi(t,r,\theta,\phi)$ along the orbit obtaining a function of 
$t$ only, 
\begin{equation}
\Psi(t)\equiv \psi(t,r_0,\frac{\pi}{2},0)=\sum_{lm} \int \frac{d\omega}{2\pi} 
Y_{lm}(\frac{\pi}{2},0) T_{lm\omega}(t),
\label{(6.6bis)}
\end{equation}
and the associated spectrum can be studied by Fourier-transforming 
with respect to time
\begin{equation}
\hat \Psi(\Omega)=\int_{-\infty}^\infty dt e^{i\Omega t} \Psi(t).
\label{(6.15)}
\end{equation}
For this quantity it is interesting to evaluate the \lq\lq soft limit" ($\Omega \to 0$) 
\begin{equation}
\lim_{\Omega \to 0}\hat \Psi(\Omega) 
=\sum_{lm}Y_{lm}(\frac{\pi}{2},0) \int_{-\infty}^{\infty} dt 
\int \frac{d\omega}{2\pi}  T_{lm\omega}(t),
\label{(6.16)}
\end{equation}
which leads to a gauge-invariant information. Unfortunately, even the study of this simplified 
quantity turns out to be rather involved, and can only be performed numerically.

\end{widetext}

\subsection{Particle at rest in the $(T,\rho)$ coordinates}
\label{subs_part_at_rest}

In this case a particle at rest, with four-velocity $U=\partial_T$, is accelerated.  
The parametric equations are
\begin{equation}
T(\tau)= \tau , \qquad \rho(\tau)=\rho_0,\qquad \theta_0=\frac{\pi}{2} , \qquad \phi_0=0,
\label{(6.17)}
\end{equation}
for a particle placed on the equatorial plane $\frac{\pi}{2}$, as before.
This implies
\begin{eqnarray}
\rho_{s}&=& -4\pi q_s \delta(\rho-\rho_0)\delta(\theta-\frac{\pi}{2})\delta(\phi)
\nonumber\\
&=& -4\pi q_s \delta(\rho-\rho_0) \sum_{lm}Y_{lm}(\theta,\phi)Y^*_{lm}(\frac{\pi}{2},0),
\nonumber\\
\label{(6.18)}
\end{eqnarray}
where now $\sqrt{-g}\big|_{\rm orbit}=1$.
Equation \eqref{scalar_field_eq_source} then becomes
\begin{equation}
\label{scalar_field_eq_2}
\left(\Box  -\mu^2\right)  \psi=-4 \pi  q_s  \delta(\rho-\rho_0) 
\sum_{lm}Y_{lm}(\theta,\phi)Y^*_{lm}(\frac{\pi}{2},0) ,
\end{equation}
and we look for solutions of the type
\begin{equation}
\psi(T,\rho,\theta,\phi)=\sum_{lm} \int \frac{d\omega}{2\pi}e^{-i\omega T} 
Y_{lm}(\theta,\phi) R_{lm\omega}(\rho).
\label{(6.20)}
\end{equation}
We are left with the following radial equation:
\begin{equation}
{\mathcal D}_\rho R_{lm\omega}=S_{lm}\delta(\rho-\rho_0),
\label{(6.21)}
\end{equation}
where
\begin{eqnarray}
\label{eq_fin_in_rho}
{\mathcal D}_\rho &=&(1-\rho^2)\frac{d^2}{d\rho^2}-2\rho\frac{d }{d\rho }
-\left({\mathcal L}-\frac{\omega^2}{(1-\rho^2)}\right)
\nonumber\\
&=& \frac{d}{d\rho}(1-\rho^2)\frac{d}{d\rho}-\left({\mathcal L}
-\frac{\omega^2}{(1-\rho^2)}\right),
\nonumber\\
S_{lm}&=&-4 \pi  q_s Y^*_{lm}(\frac{\pi}{2},0) .
\label{(6.22)}
\end{eqnarray}
It is convenient to introduce a tortoise-like coordinate
\begin{equation}
\rho_*={\rm arctanh}(\rho),
\label{(6.23)}
\end{equation}
such that $\rho\in (-1,1)$ is mapped into $\rho_*\in (-\infty,\infty)$ and
\begin{equation}
\frac{d^2}{d\rho_*^2}=(1-\rho^2)\left[(1-\rho^2)\frac{d^2}{d\rho^2}-2\rho\frac{d }{d\rho }\right].
\label{(6.24)}
\end{equation}
Therefore, Eq. \eqref{(6.21)} can be written as
\begin{equation}
\left[\frac{d^2}{d\rho_*^2}+\left(\omega^2 -V_{\rm PT}(\rho_*)\right)\right] 
R_{lm\omega}=C_{lm}\delta(\rho-\rho_0),
\label{(6.25)}
\end{equation}
with
\begin{equation}
C_{lm}=\frac{S_{lm}}{\cosh^2 \rho_0}\,,
\label{(6.26)}
\end{equation}
and
\beq
V_{\rm PT}(\rho_*)=\frac{{\mathcal L}}{\cosh^2\rho_*}\,,
\eeq
the well known P\"oschl-Teller potential \cite{Wardell}.
We can write
\begin{equation}
{\mathcal L}=-\lambda (\lambda +1),
\label{(6.27)}
\end{equation}
($\lambda$ is a function of $l$ since ${\mathcal L}$  is defined in Eq. \eqref{(5.7)}, 
 but it should not be confused with $l$ and $\widehat l$)
with inverse relation
\begin{equation}
\lambda=-\frac12 \pm \frac12 \sqrt{1-4{\mathcal L}}\,,
\label{(6.28)}
\end{equation}
As stated above, Eq. \eqref{(6.25)} with ${\mathcal L}$ 
given in Eq. \eqref{(6.27)}  can be solved in terms of Legendre functions
\begin{equation}
R_{lm\omega}(\rho)=C_1 P_{\lambda, i\omega}(\rho)+C_2 Q_{\lambda, i\omega}(\rho),
\label{(6.29)}
\end{equation}
with $P_{\lambda, i\omega}(\rho)$ and $Q_{\lambda, i\omega}(\rho)$ defined in Eqs. 
\eqref{PQ_defs} in terms of hypergeometric functions. Note that ${\mathcal L}$ is a 
function of $l$ and hence $\lambda$ is also a function of $l$: while denoting 
$P_{\lambda, i\omega}$ and $Q_{\lambda, i\omega}$ with an explicit dependence on 
$\lambda$, in the combination $R_{lm\omega}$ of Eq. \eqref{(6.29)} we restore the dependence on $l$.
We need two linearly independent solutions of Eq. \eqref{(6.25)} with $\rho\in (-1,1)$.
The observation that Eq. \eqref{(6.25)} depends on $\lambda$ via 
the combination $\lambda(\lambda+1)$, which 
is invariant if $\lambda \to -\lambda-1$, implies that, 
if $P_{\lambda, i\omega}(\rho)$ is a solution,
the same holds for $P_{-\lambda-1, i\omega}(\rho)$, i.e.,
also this replacement leads to linearly 
independent solutions. The same is true for $\omega\to-\omega$ and 
$\rho\to -\rho$ or $\rho_*\to-\rho_*$.
Following Ref. \cite{Wardell} let us choose as independent solutions  
$P_{\lambda, i\omega}(\rho)\equiv R_{lm\omega}^{\rm up}(\rho)$ (regular at $\rho=1$) 
and $P_{\lambda, i\omega}(-\rho)\equiv R_{lm\omega}^{\rm in}(\rho)$ 
(regular at $\rho=-1$), exploiting here the nomenclature \lq\lq in" and \lq\lq up" 
of black hole perturbation theory. Their constant Wronskian is
\begin{eqnarray}
W_{\lambda,\omega}&=&(1-\rho^2)(R_{lm\omega}^{\rm in}R_{lm\omega}^{\rm up}{}'
-R_{lm\omega}^{\rm up}R_{lm\omega}^{\rm in}{}')
\nonumber\\
&=& \frac{2i\omega\Gamma(-i\omega)}{\Gamma(1-i\omega) } 
\frac{1}{\Gamma(l+1-i\omega)\Gamma(-l-i\omega)}.\qquad
\label{(6.30)}
\end{eqnarray}
We then have the following Green function for Eq. \eqref{(6.25)}:
\begin{eqnarray}
G_{\lambda \omega}(\rho,\rho')&=&\frac{1}{W_{\lambda,\omega}}
\left[R_{lm\omega}^{\rm in}(\rho)R_{lm\omega}^{\rm up}(\rho') H(\rho'-\rho)\right.
\nonumber\\
&+& \left. R_{lm\omega}^{\rm in}(\rho')R_{lm\omega}^{\rm up}(\rho ) H(\rho -\rho')
\right]\nonumber\\
&\equiv & G_{\lambda \omega}^-(\rho,\rho')  H(\rho'-\rho)\nonumber\\
&+&G_{\lambda \omega}^-(\rho',\rho)  H(\rho-\rho')\,,
\label{(6.31)}
\end{eqnarray}
leading to a left and a right $R_{lm\omega}(\rho )=S_{lm}G_{\lambda \omega}(\rho,\rho_0)$, 
i.e., we get for the left and right fields
\begin{eqnarray}
R_{lm\omega}^-(\rho )&=& \frac{S_{lm}}{W_{\lambda,\omega}} 
R_{lm\omega}^{\rm in}(\rho) R_{lm\omega}^{\rm up}(\rho_0), 
\nonumber\\
R_{lm\omega}^+(\rho )&=& \frac{S_{lm}}{W_{\lambda,\omega}} 
R_{lm\omega}^{\rm up}(\rho) R_{lm\omega}^{\rm in}(\rho_0).
\label{(6.32)}
\end{eqnarray}
By virtue of the continuity of the field, we can limit our considerations, 
for example, to the left field. Hence we find 
\begin{eqnarray}
\label{psi_min}
\psi^-(T,\rho,\theta,\phi)&=& \sum_{lm}Y_{lm}S_{lm}\int \frac{d\omega}{2\pi}
e^{-i\omega T}G_{\lambda \omega}^-(\rho,\rho_0)\,,
\nonumber\\
\end{eqnarray}
We can now evaluate $\psi^-(T,\rho,\theta,\phi)=\psi(T,\rho,\theta,\phi)$  
along the particle's world line
\begin{eqnarray}
\psi^-(T,\rho_0,\frac{\pi}{2},0)&=& \sum_{l}Y_{l0}(\frac{\pi}{2},0)S_{l0} 
\nonumber\\
&& \times \int \frac{d\omega}{2\pi}e^{-i\omega T}G_{\lambda \omega}^-(\rho_0,\rho_0).\qquad
\label{(6.35)}
\end{eqnarray}
Even taking $\rho_0\ll 1$ and expanding the hypergeometric functions for small 
$\rho$, the Fourier transform can be only performed numerically, since $\omega$ 
remains inside the Gamma functions.
We limit therefore ourselves to expressing the result in the $\omega$ space
\begin{eqnarray}
\hat \psi^-(\omega,\rho_0,\frac{\pi}{2},0)&=& 
\sum_{l}Y_{l0}(\frac{\pi}{2},0)S_{l0}G_{\lambda \omega}^-(\rho_0,\rho_0) ,\qquad
\label{(6.36)}
\end{eqnarray}
where $\hat \psi^-(\omega,\rho_0,\frac{\pi}{2},0)$ denotes the Fourier transform 
of $\psi^-(T,\rho_0,\frac{\pi}{2},0)$ with respect to the variable $T$.
The Fourier domain spectrum of the field along the source's world line in 
this case is represented by the function $G_{\lambda \omega}^-(\rho,\rho_0)$  
which 
can be studied in the soft limit $\omega \to 0$. We will not continue along 
this path because of the difficulties in obtaining analytic results.

Eventually, let us notice that the asymptotic behaviors 
\begin{eqnarray}
\lim_{\rho_*\to -\infty }R_{lm\omega}^{\rm in}(\rho)&\sim& e^{-i\omega \rho_*},
\nonumber\\
\lim_{\rho_*\to +\infty }R_{lm\omega}^{\rm up}(\rho) &\sim&  e^{+i\omega \rho_*},
\label{(6.37)}
\end{eqnarray}
are automatically taken into account.

\section{Scalar self-force and its variations}

On denoting by $u^\alpha$ the source's four-velocity, the definition 
of scalar self-force along the source's world line reads as
\begin{equation}
F_\alpha(\tau) =q P(u)_\alpha{}^\beta \partial_\beta \psi \bigg|_{x^\alpha=x^\alpha(\tau)}
\label{(7.1)}
\end{equation}
where $P(u)_\alpha{}^\beta=\delta_\alpha^\beta +u_\alpha u^\beta$ 
projects orthogonally onto $u^\alpha$.

In general, one has to distinguish between left/right force components, 
$F_\alpha^\pm(\tau)$ when approaching the (singular) worldline of the source 
from left or from right. In fact, unlike the field which is continuous 
across the source worldline, its derivatives are in general discontinuous.
Let us work, for example, with the Nariai metric written in dS-like 
coordinates $(T,\rho,\theta,\phi)$ and 
with the source at rest with respect to the coordinates as discussed 
in Subsection \ref{subs_part_at_rest}.

The field at a generic spacetime point has been given above (see e.g., 
Eq. \eqref{psi_min}) and rewritten below for convenience
\begin{eqnarray}
\psi^\pm (T,\rho,\theta,\phi)&=& \sum_{lm}Y_{lm}S_{lm}\int \frac{d\omega}
{2\pi}e^{-i\omega T}G^\pm _{lm\omega}(\rho,\rho_0),
\nonumber\\
\label{(7.2)}
\end{eqnarray}
where
\begin{eqnarray}
G^-_{lm\omega}(\rho,\rho_0)&=&\frac{R_{lm\omega}^{\rm in}(\rho)
R_{lm\omega}^{\rm up}(\rho_0)}
{W_{\lambda,\omega}},
\nonumber\\
G^+_{lm\omega}(\rho,\rho_0)&=&\frac{R_{lm\omega}^{\rm in}(\rho_0)
R_{lm\omega}^{\rm up}(\rho)}{W_{\lambda,\omega}}.
\label{(7.3)}
\end{eqnarray}
Therefore, in general,
\begin{eqnarray}
\psi^\pm (T,\rho,\theta,\phi)&=& \sum_{lm}Y_{lm}\int 
\frac{d\omega}{2\pi}e^{-i\omega T}R_{lm\omega}^\pm(\rho)\,,
\label{(7.4)}
\end{eqnarray}
with $R_{lm\omega}^\pm(\rho)=S_{lm}G^\pm_{lm\omega}(\rho,\rho_0)$, as before.
For example when $u^\alpha=u^T\delta^\alpha_T$ with $u^T=(1-\rho^2)^{-1/2}$, 
the orthogonality condition $F_\alpha u^\alpha=0$  implies that $F_T=0$, and hence
\begin{eqnarray}
F_\rho 
&=&q \partial_\rho \psi ,
\nonumber\\
F_\theta 
&=&  q  \partial_\theta \psi ,
\nonumber\\
F_\phi &=& q  \partial_\phi \psi,
\label{(7.5)}
\end{eqnarray}
where, as already stated, now we have to consider both the left and 
right parts of the field separately
\begin{equation}
\psi^\pm(T,\rho,\theta,\phi)=\sum_{lm}Y_{lm}(\theta,\phi)\int 
\frac{d\omega}{2\pi}e^{-i\omega T}   R_{lm\omega}^\pm(\rho),
\label{(7.6)}
\end{equation}
with $\psi^+(T,\rho,\theta,\phi)=\psi^-(T,\rho,\theta,\phi)$ but
\begin{eqnarray}
\partial_T \psi^\pm&=&\sum_{lm}Y_{lm}(\theta,\phi)\int \frac{d\omega}
{2\pi}e^{-i\omega T}  (-i\omega) R_{lm\omega}^\pm(\rho),
\nonumber\\
\partial_\rho \psi^\pm&=&\sum_{lm}Y_{lm}(\theta,\phi)\int 
\frac{d\omega}{2\pi}e^{-i\omega T}\partial_\rho R_{lm\omega}^\pm(\rho),
\nonumber\\
\partial_\phi \psi^\pm&=&\sum_{lm}(im) Y_{lm}(\theta,\phi)\int 
\frac{d\omega}{2\pi}e^{-i\omega T}  R_{lm\omega}^\pm(\rho).
\nonumber\\
\label{(7.7)}
\end{eqnarray}
This requires the definition of a corresponding left and right force, 
$F_\alpha^\pm$, evaluated along the orbit, using
$\partial_T \psi^\pm(T,\rho,\theta,\phi)\big|_{\rm orb}$,   
$\partial_\phi \psi^\pm(T,\rho,\theta,\phi)\big|_{\rm orb}$, 
$\partial_\rho \psi^\pm(T,\rho,\theta,\phi)\big|_{\rm orb}$, 
which becomes a function of the parameter used along the orbit.
Indeed, once the derivatives of $\psi$ have been taken and restricted to 
the orbit, one can define the left and right components of the force 
(along the particle's world line),
\begin{eqnarray}
F_\rho^\pm,\quad
F_\theta^\pm ,\quad
F_\phi^\pm.
\label{(7.8)}
\end{eqnarray}
All this machinery can be easily activated, but the 
main tool for analyzing results is, so far, numerical.

\subsection{Can a particle \lq\lq feel" the presence of the superimposed test field?}

A different, non-standard approach can be that of defining a force from 
the energy-momentum tensor associated with a scalar field,
\begin{equation}
T_{\mu\nu}= (\nabla_\mu \psi) (\nabla_\nu \psi) -\frac12 g_{\mu\nu} 
(\nabla_\alpha \psi) (\nabla^\alpha \psi),
\label{(7.9)}
\end{equation}
computable by using the above relations.

In principle, all perturbations which do not modify the background 
spacetime do not interact directly among themselves 
(while each of them interacts with the background, by making  
use of the covariant derivative when defining the perturbation itself).  
At the next level, one has to consider backreaction of the field on 
the background (so far, the field is no longer definable a \lq\lq test" 
field). Often this is not a simple task, and if one wants to have a 
description for a particle moving on the Nariai spacetime and which {\it also} 
interacts with the perturbing field, the interaction should be defined.
A possibility is to use the so-called (generalized) Poynting-Robertson (PR) approach.
This PR scheme  
\cite{Poy,Robertson,WW,Bini:2008vk,Bini:2010xa,Bini:2011cll,Bini:2012kb,Bini:2012ncd,EB1,EB2}
has been mainly studied in black hole spacetimes.

An external force is introduced which is responsible for deviating  
particle motions with four-velocity $u$ from their geodesic behavior:
\begin{equation}
\label{F_ext}
{\mathcal F}_{\rm ext}(u)^\alpha= \sigma P(u)^\alpha{}_\beta T^{\beta\nu} u_\nu ,
\end{equation}
where $\sigma$ is a parameter needed for dimensional reasons, $\sigma \sim L$, 
and to model the intensity of the force itself~\footnote{We will not 
specify the sign of $\sigma$ since the energy-momentum tensor of the field might 
act either as generating a friction force on the particle or, being attracted 
by the underlying spacetime gravity effects as the particle itself, might 
contribute with an additional extra acceleration. We will not discuss the 
details of this approach, which among other things was originally developed 
with a field of photons and not for a scalar field or a fluid.} and 
$T^{\mu\nu}$ is the energy-momentum tensor of the superimposed test fluid field.
It is worth remarking the $u^\alpha$-structure of ${\mathcal F}_{\rm ext}(u)^\alpha$ which 
is odd (cubic) in $u^\alpha$, i.e.,
\begin{equation}
{\mathcal F}_{\rm ext}(u)^\alpha\sim A^\alpha{}_{\beta\gamma\delta} 
u^\beta u^\gamma u^\delta +B^\alpha{}_{\beta} u^\beta,
\end{equation}
a good candidate for a covariant definition of a friction force.

A formally Newtonian force equation is then studied
\begin{equation}
m a(u)^\alpha= {\mathcal F}_{\rm ext}(u)^\alpha,
\label{(7.11)}
\end{equation}
where $a(u)^\alpha=\nabla_u u^\alpha$ is the four-acceleration of the particle.
In the definition of ${\mathcal F}_{\rm ext}(u)^\alpha$ it is of fundamental 
importance to satisfy the conservation law $T^{\beta\nu}{}_{;\nu}=0$, 
while the presence of the projector orthogonal to $u$, $P(u)$, is needed 
for consistency, because the acceleration $a(u)$ is orthogonal to $u$ by definition.
Having said this, ${\mathcal F}_{\rm ext}(u)$ should be nothing but a useful 
approximation, to be adopted, for example, when the full perturbation theory approach 
is rather involved, as is often the case.

In this  case  Eqs. \eqref{F_ext} and \eqref{(7.9)} imply
\begin{eqnarray}
\label{F_ext2}
{\mathcal F}_{\rm ext}(u)^\alpha &=& \sigma P(u)^\alpha{}_\beta 
\Bigr[(\nabla^\beta  \psi) (\nabla^\nu \psi) \nonumber \\ 
& \; & -\frac12 g^{\beta\nu} (\nabla_\rho \psi) (\nabla^\rho \psi)\Bigr] u_\nu
\nonumber\\
&=& \sigma (\nabla_u \psi) P(u)^\alpha{}_\beta  (\nabla^\beta \psi). 
\end{eqnarray}
Comparing ${\mathcal F}_{\rm ext}(u)$ with the definition of self-force $F$ given 
in Eq. \eqref{(7.1)} we find that these two forces are proportional, i.e.,
\beq
{\mathcal F}_{\rm ext}(u)^\alpha= \frac{\sigma}{q} (\nabla_u \psi) F^\alpha\,,
\eeq
an interesting feature that we will explore in future works.

\section{Can Nariai spacetime support a test fluid?}
\label{test_fluid}

\begin{figure}
\includegraphics[scale=0.35]{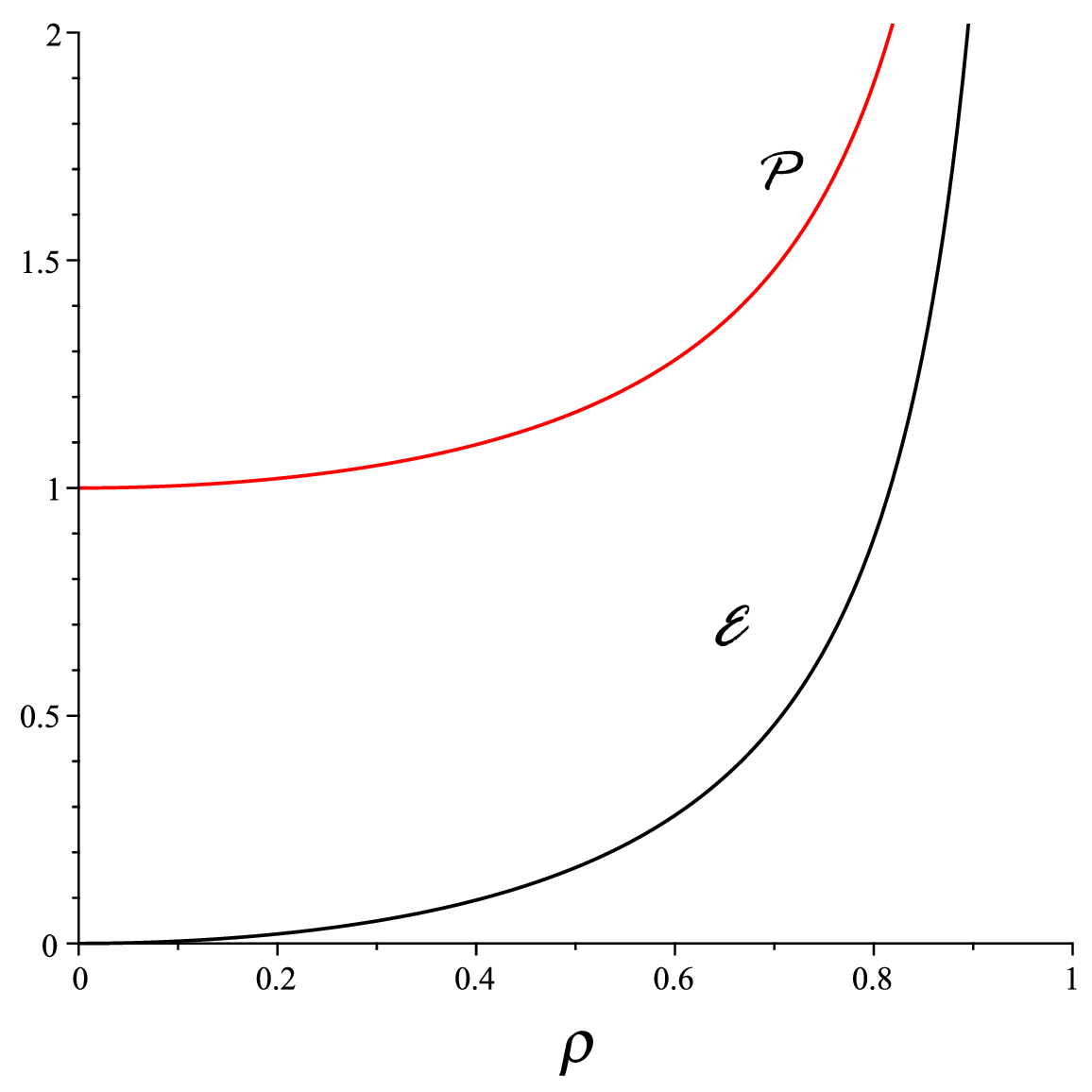}
\caption{\label{fig:en_pres_iso} Behavior of energy density and pressure for the test fluid 
of Eq. \eqref{t_en_im} and $U$ having constant frame components as in Eq. \eqref{(4.17)}. 
The parameters here have been chosen so that $A(0)=1=B(0)$, $v=0.5$.
} 
\end{figure}

Let us consider the metric written in coordinates $(T,\rho,\theta,\phi)$ and an 
observer moving radially (with velocity $v$) with respect to the coordinates. 
An example is provided by the observer $U$ defined in Eq. \eqref{(4.17)} above, having constant 
frame components for the four-velocity (see Eqs. \eqref{(4.17)} and \eqref{(4.18)}).
This observer exists in the region $ \rho^2 \le 1$ and can be taken as identifying 
the rest frame of a test fluid with the energy-momentum tensor
\begin{equation}
\label{t_en_im}
T_{\alpha\beta}=A(\rho)U_\alpha U_\beta +B(\rho) g_{\alpha\beta},
\end{equation}
where
\begin{equation}
A(\rho)={\mathcal E}(\rho)+{\mathcal P}(\rho) , \qquad B(\rho)={\mathcal P}(\rho),
\label{(8.2)}
\end{equation}
in terms of energy density and pressure.
The conservation law $T_{\alpha\beta}{}^{;\beta}=0$ imposed by considering this field 
superimposed on the Nariai spacetime leads to
\begin{eqnarray}
\label{AB_iso}
A(\rho) &=& \frac{A(0)}{(1-\rho^2)},
\nonumber\\
B(\rho) &=&  \frac{A(0)}{2(1-\rho^2)}+B(0)-\frac12 A(0),
\end{eqnarray}
implying that ${\mathcal P}(\rho)=B(\rho)$ and
\begin{equation}
{\mathcal E}(\rho)= \frac{(2-\rho^2)}{2(1-\rho^2)}A(0)-B(0).
\label{(8.4)}
\end{equation}
The behaviors of energy and pressure are shown in Fig. \ref{fig:en_pres_iso}.
Interestingly, the case $A(0)=2$, $B(0)=1$ is peculiar in the sense that, for these values of 
$A(0)$ and $B(0)$, one finds ${\mathcal E}(\rho)={\mathcal P}(\rho)$. 

An important variant of this example consists in using an observer still 
moving radially but with anisotropic velocity. For example,
\begin{equation}
U_{\rm anis}=\frac{1}{\sqrt{1-\rho^2-v^2}} \left[\partial_T 
+ v\sqrt{1-\rho^2}\, \partial_\rho\right].
\label{(8.5)}
\end{equation}
The physical velocity for this observer is obtained by restoring the frame vectors
\begin{equation}
U_{\rm anis}=\gamma_{\rm phys} \left[ e_0 
+ v_{\rm phys} e_1\right],
\label{(8.6)}
\end{equation}
where
\begin{equation}
v_{\rm phys}=\frac{v}{\sqrt{1-\rho^2}},\qquad \gamma_{\rm phys}=\frac{1}{\sqrt{1-v_{\rm phys}^2}}\,,
\label{(8.7)}
\end{equation}
which, depending on $\rho$, marks a substantial difference with  
respect to the observer defined in Eq. \eqref{(4.17)}.

\begin{figure}
\includegraphics[scale=0.35]{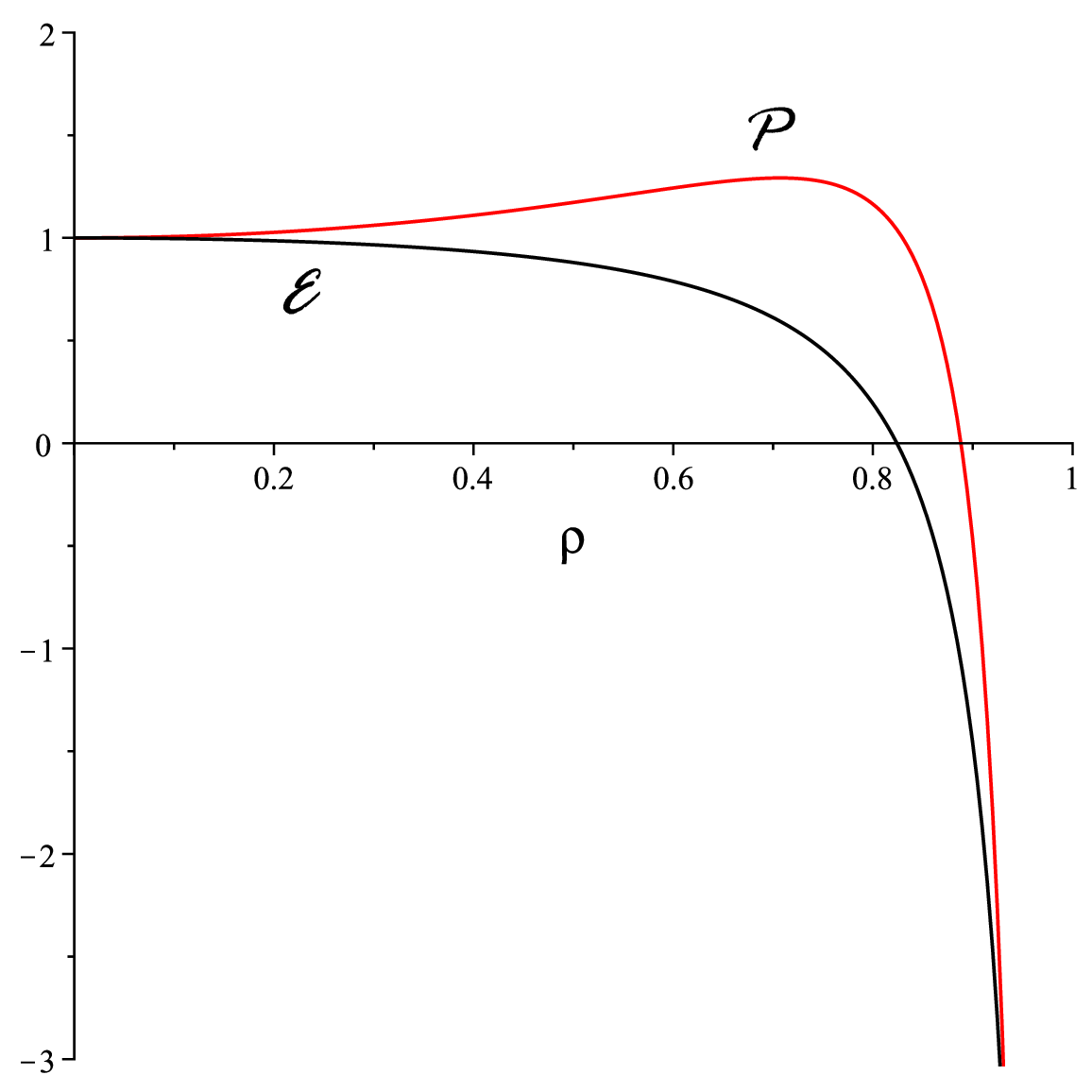}
\caption{\label{fig:en_pres} Behavior of energy density and pressure for the test fluid 
of Eq. \eqref{t_en_im}. The parameters have been chosen so that $A_{\rm anis}(0)=2$, 
$B_{\rm anis}(0)=1$, $v=0.5$. The fluid starts being exotic at $\rho=0.8241$ where 
the energy density vanishes. The pressure vanishes at $\rho=0.8882$ and reaches its 
maximum value ${\mathcal P}_{\rm max}=1.2919$ at $\rho=0.7071$. Moreover, it has its 
relative minimum value ${\mathcal P}_{\rm min}=1.2919$ at $\rho=0$. The energy density 
is instead monotonically decreasing from the value at $\rho=0$.}
\end{figure}

This radially-anisotropically moving observer exists in the 
region $ \rho^2 \le 1-v^2$ and can 
be taken again as identifying the rest frame of a test fluid, with the 
energy-momentum tensor \eqref{t_en_im} defined above. In this case 
\begin{eqnarray}
A_{\rm anis}(\rho)&=& \frac{1-\gamma^2\rho^2}{(1-\rho^2)^{3/2}}A_{\rm anis}(0),
\nonumber\\
B_{\rm anis}(\rho)&=& \frac{\gamma^2}{3} 
\left[\frac{[3(1-\rho^2)-2v^2]}{(1-\rho^2)^{3/2}}
- (3-2v^2)\right]A_{\rm anis}(0)
\nonumber\\
&+&B_{\rm anis}(0),
\nonumber\\
\label{(8.8)}
\end{eqnarray}
where $\gamma=(1-v^2)^{-1/2}$, and the constants $A_{\rm anis}(0)$ and 
$B_{\rm anis}(0)$ can be chosen arbitrarily. In particular, 
$B_{\rm anis}(\rho)={\mathcal P}(\rho)$ and
\begin{eqnarray}
{\mathcal E}(\rho)&=&\frac{A_{\rm anis}(0)\gamma^2}{3(1-\rho^2)^{3/2}}
[-v^2+(3-2v^2)(1-\rho^2)^{3/2}]
\nonumber\\
&-& B_{\rm anis}(0).
\label{(8.9)}
\end{eqnarray}
For $v\in [0,1)$ there exists always a physically allowed region for the fluid.
In the special case $v=0$ the above solution takes the simpler form
\begin{eqnarray}
{\mathcal E}(\rho)&=& A_{\rm anis}(0) -B_{\rm anis}(0)={\rm const.},
\nonumber\\
{\mathcal P}(\rho)&=&  \left[\frac{1}{(1-\rho^2)^{1/2}}-1\right]A_{\rm anis}(0)
\nonumber\\
&+& B_{\rm anis}(0).
\label{(8.10)}
\end{eqnarray}
The behaviors of energy and pressure in this radially anisotropic case are 
shown in Fig. \ref{fig:en_pres}.

In order to provide an explicit example let us consider $u$ moving radially 
(as the fluid), see Eq. \eqref{(4.17)} above,
\begin{equation}
u=\gamma_V (e_0+V e_1).
\label{(8.11)}
\end{equation}
Thus
\begin{eqnarray}
T^{\beta\nu} u_\nu &=&[A(\rho) U^\beta U^\nu+B(\rho) g^{\beta\nu}]u_\nu
\nonumber\\
&=& A(\rho) U^\beta (U\cdot u)+B(\rho) u^\beta\,,
\label{(8.12)}
\end{eqnarray}
since
\begin{equation}
U=\gamma (e_0+ve_1).
\label{(8.13)}
\end{equation}
Therefore
\begin{eqnarray}
{\mathcal F}_{\rm ext}(u)^\alpha &=& \sigma P(u)^\alpha{}_\beta [A(\rho) 
U^\beta (U\cdot u)+B(\rho) u^\beta]
\nonumber\\
&=& \sigma A(\rho)  P(u)^\alpha{}_\beta  U^\beta (U\cdot u)
\nonumber\\
&=& \sigma A(\rho) \gamma_V^2(v-V)E_1^\alpha ,
\label{(8.14)}
\end{eqnarray}
having used $E_1=\gamma[ve_0+  e_1]$ ($U\cdot E_1=0$) and
\begin{equation}
U\cdot u=\gamma \gamma_V(vV-1).
\label{(8.15)}
\end{equation}
Therefore, from Eqs. \eqref{(4.18)} and \eqref{AB_iso}, the motion is 
ruled by the equation
\begin{equation}
-m\frac{\gamma \rho}{\sqrt{1-\rho^2}}=\sigma \frac{A(0)}{(1-\rho^2)} \gamma_V^2(v-V).
\label{(8.16)}
\end{equation}
This means that the particle is \lq\lq suspended", i.e., it 
can stay at a fixed $\rho$ such that
\begin{equation}
\rho\sqrt{1-\rho^2} =-\frac{\sigma}{m} A(0)\frac{\gamma_V^2(v-V) }{\gamma } ,
\label{(8.17)}
\end{equation}
which can be satisfied for various choices of the parameters (see e.g., \cite{Bini:2008vk}). 
In this (very special and simple) case the acceleration of the particle is 
compensated by the external force. The special case $\rho=0$ corresponds to the 
trivial situation $v=V$.

One can push forward the present analysis by considering the most general situation 
of a particle (with mass $m$ and worldline parametrized by the proper time 
$\tau$) moving on the equatorial plane ($\theta=\frac{\pi}{2}$) and undergoing 
acceleration effects due to the external force
${\mathcal F}_{\rm ext}$, \eqref{F_ext}, generated by the test fluid of Eqs. 
\eqref{t_en_im}, \eqref{(8.2)} and \eqref{AB_iso}.
The unit timelike four-velocity of the particle is given by
\begin{equation}
\label{u_equat}
u=\frac{dT}{d\tau}\partial_T+\frac{d\rho}{d\tau}\partial_\rho
+\frac{d\phi}{d\tau}\partial_\phi ,  
\end{equation}
and its motion is described by the equations
\begin{equation}
a(u)^\alpha=\nabla_u u^\alpha =\tilde \sigma\, P(u)^\alpha{}_\beta T^{\beta\nu}u_\nu,
\label{new_eqs_motion}
\end{equation}
where
\begin{equation}
\tilde \sigma=\frac{\sigma}{m}.
\end{equation}
Explicitly,
\begin{eqnarray}
\label{aU_compts}
a(u)^T &=&\frac{du^T}{d\tau}-\frac{2\rho}{(1-\rho^2)}u^T u^\rho, 
\nonumber\\
a(u)^\rho &=&\frac{du^\rho}{d\tau}+\frac{\rho}{(1-\rho^2)}[(u^\rho)^2-(1-\rho^2)^2(u^T)^2], 
\nonumber\\
a(u)^\theta &=&0, 
\nonumber\\
a(u)^\phi &=&\frac{du^\phi}{d\tau},
\end{eqnarray}
with
\begin{equation}
\label{eq_normaliz}
u^\rho=\pm \sqrt{1-\rho^2} \left[(1-\rho^2)(u^T)^2-(u^\phi)^2-1 \right]^{1/2}.
\end{equation}

The equations of motion depend on several parameters: the fluid parameters $A(0)$, 
$B(0)$ and the quantity $\tilde \sigma=\sigma/m$, used to build a model of particle-fluid interaction.
The limiting case $\tilde \sigma=0$ corresponds to geodesic motion, 
whereas the situations $\tilde \sigma >0$ and $\tilde \sigma <0$ imply an additional 
positively accelerated motion (a sort of \lq\lq favorable wind") and an additional 
negatively accelerated motion (a sort of \lq\lq headwind", or a friction force). 
Moreover, the radial equation can be either integrated inward or forward, i.e., one 
can freely choose the sign in front of $\frac{d\rho}{d\tau}$ in Eq. \eqref{eq_normaliz}, 
obtained from the timelike condition $u\cdot u=-1$.
Actually, we can measure deviations from geodesic motion 
in terms of $\tilde \sigma$. We have studied 
(mainly numerically) the equations \eqref{new_eqs_motion}, assuming $A(0)=1=B(0)$ for simplicity.

In this special case, we find
\begin{eqnarray}
\label{F_PR_plots}
{\mathcal F}_{\rm ext}^T&=&\sigma\left[ q_{T\phi\phi}^T u^T(u^\phi)^2
+ q_{T\rho\rho}^T u^T(u^\rho)^2+q_{TT\rho}^T(u^T)^2 u^\rho\right.
\nonumber\\
&+&\left. q_{TTT}^T (u^T)^3+ q_\rho^T u^\rho+q_T^T u^T\right],
\nonumber\\
{\mathcal F}_{\rm ext}^\rho&=&\sigma \left[ q_{\rho\phi\phi}^\rho u^\rho(u^\phi)^2
+ q_{T\rho\rho}^\rho u^T(u^\rho)^2
+q_{TT\rho}^\rho (u^T)^2 u^\rho\right.
\nonumber\\
&+&\left. q_{\rho\rho\rho}^\rho (u^\rho)^3+ q_\rho^\rho u^\rho+q_T^\rho u^T\right],
\nonumber\\
{\mathcal F}_{\rm ext}^\theta&=&0,
\nonumber\\
{\mathcal F}_{\rm ext}^\phi&=&\sigma \left[ q_{\phi\phi\phi}^\phi  (u^\phi)^3
+ q_{\phi\rho\rho}^\phi u^\phi(u^\rho)^2
+q_{\phi T\rho}^\phi  u^T u^\phi u^\rho\right.
\nonumber\\
&+&\left. q_{TT\phi}^\phi (u^T)^2 u^\phi+ q_\rho^\phi u^\rho \right],
\end{eqnarray}
with
\begin{eqnarray}
q_{T\phi\phi}^T&=&q_{\rho\phi\phi}^\rho=q_{\phi\phi\phi}^\phi=q_\rho^\phi=\frac12 
\frac{2-\rho^2}{(1-\rho^2)}, 
\nonumber\\ 
q_{T\rho\rho}^T&=&q_{\rho\rho\rho}^\rho=q_{\phi\rho\rho}^\phi=-q_\rho^\rho 
= \frac12 \frac{2\gamma^2-\rho^2}{(1-\rho^2)^2},
\nonumber\\ 
q_{TT\rho}^T &=&q_{T\rho\rho}^\rho=q_{\phi T\rho}^\phi=-\frac{2\gamma^2 v}{(1-\rho^2)} ,
\nonumber\\ 
q_{TTT}^T&=&q_{TT\rho}^\rho=\frac12 (\rho^2+2\gamma^2v^2), 
\nonumber\\  
q_\rho^T&=& \frac{\gamma^2 v}{(1-\rho^2)^2},
\nonumber\\  
q_T^T&=& -\frac12 \frac{\rho^2+2\gamma^2 v^2}{(1-\rho^2)},
\nonumber\\ 
q_T^\rho&=& -\gamma^2 v,
\nonumber\\ 
q_{TT\phi}^\phi&=& \frac12 (\rho^2-\gamma^2 v^2).
\end{eqnarray}
Examples of numerical integration results are shown in Figures \ref{fig:rho_di_tau_PR}, 
\ref{fig:t_di_tau_PR} and \ref{fig:phi_di_tau_PR} below.
Analytical integration can be performed as well. In this case, however, 
it is convenient to fully specify either the geodesic orbit which is going to be 
perturbed by the Poynting-Robertson-like external force or the test fluid equation 
of state. As before we assume then $A(0)=1=B(0)$.
For the background (equatorial) geodesics we choose instead $E=\sqrt{2}$ and $L=1$, implying
\begin{eqnarray}
\label{geo_ref}
T_{\rm geo}(\tau) &=&\sqrt{2}\tau\mp\frac{1}{2}\ln \left(\frac{4 \mp 2 \sqrt{2}\tau}{3} \right) ,
\nonumber\\
\rho_{\rm geo}(\tau) &=& \frac12 e^{\pm \sqrt{2}\tau},
\nonumber\\
\phi_{\rm geo}(\tau) &=& \tau, 
\end{eqnarray}
as shown above.
Even in this quite simple case the first-order corrections in $\tilde \sigma$
\begin{eqnarray}
T(\tau)&=&T_{\rm geo}(\tau) +\tilde \sigma T_1(\tau), 
\nonumber\\
\rho(\tau)&=&\rho_{\rm geo}(\tau) +\tilde \sigma \rho_1(\tau), 
\nonumber\\
\phi(\tau)&=&\phi_{\rm geo}(\tau) +\tilde \sigma \phi_1(\tau) ,
\end{eqnarray}
are rather involved. For example, on choosing the negative sign in 
$\pm 1$ we find
\begin{eqnarray}
\phi_1(\tau)&=&\frac{\pi^2}{12}+ \Bigr(\ln(2) + \sqrt{2}\tau - \frac34 \Bigr)
\ln(2e^{\sqrt{2}\tau} - 1) 
\nonumber\\
&-& \frac{1}{12}\ln(2e^{\sqrt{2}\tau} + 1) 
+ \frac{1}{12}\ln(3) 
+ C_5\tau 
+ \frac{5\sqrt{2}\tau}{6} 
\nonumber\\
&-&\frac13 {\rm dilog}(2e^{\sqrt{2}\tau} + 1) 
+ \frac13 {\rm dilog}(3)  
+ {\rm dilog}(2e^{\sqrt{2}\tau})
\nonumber\\
\end{eqnarray}
(the sector $T-\rho$ is  decoupled from the $\theta-\phi$ one also in this case)
with ${\rm dilog}(3)\approx -1.4367$ and having fixed integration constants in such a way that 
$\lim_{\tau\to 0}\phi_1(\tau)=0$ (with the constant $C_5$ freely specifiable until one  
also constrains $\lim_{\tau\to 0}\frac{d\phi_1}{d\tau}$). The equations for $T_1(\tau)$ and $\rho_1(\tau)$ 
can be only reduced to quadratures (leading to lengthy expressions involving both log and 
dilog functions to be integrated over $\tau$: these expressions can be anyway derived 
straightforwardly and we will not display them below), strengthening the importance of 
numerical integration as the most convenient tool to analyze these results.

\begin{figure}
\includegraphics[scale=0.35]{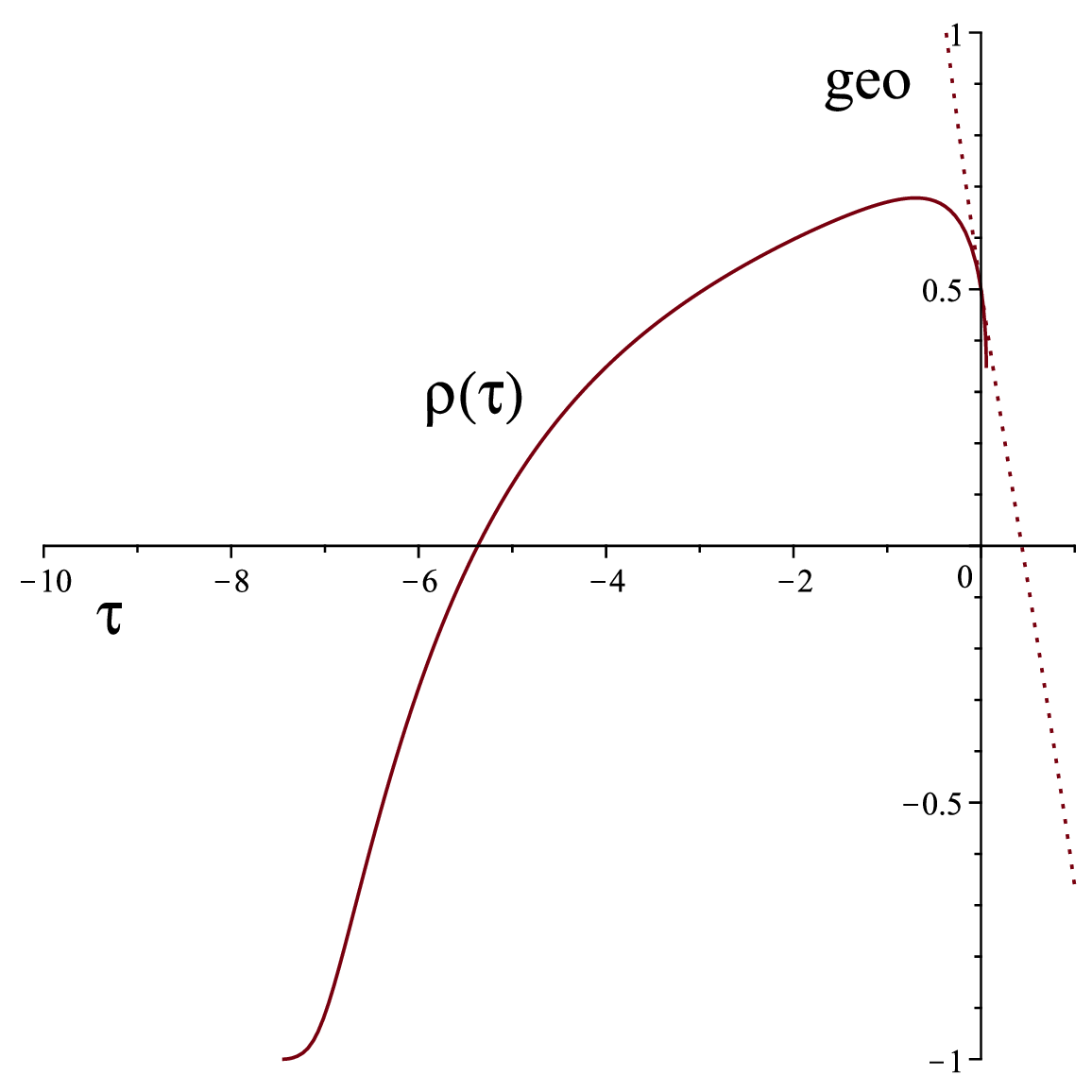}
\caption{\label{fig:rho_di_tau_PR} Comparisons of radial motions in presence of a PR external force 
as described by Eqs. \eqref{F_PR_plots} and for a geodesic. Initial conditions are chosen as follows: 
$\phi(0) = 0$, $\rho(0) = 0.5$, $t(0) = 0$,
$D(\phi)(0) = 0.1$,  $D(t)(0) = 2$ and $D(\rho)(0) = -1.2217$ (the latter condition 
derived from the normalization of $u$; the minus sign corresponds to integration inward). 
Here, we have used $v=0.5$ and $\tilde \sigma=1$ to enhance the difference with respect to the 
geodesic orbit ($\tilde \sigma=0$, dotted curve). The fluid parameters are fixed as 
$A(0)=1=B(0)$. Because of the stiffness (and the numerical precision) the plot 
cannot be displayed for positive values of $\tau$.
 }
\end{figure}

\begin{figure}
\includegraphics[scale=0.35]{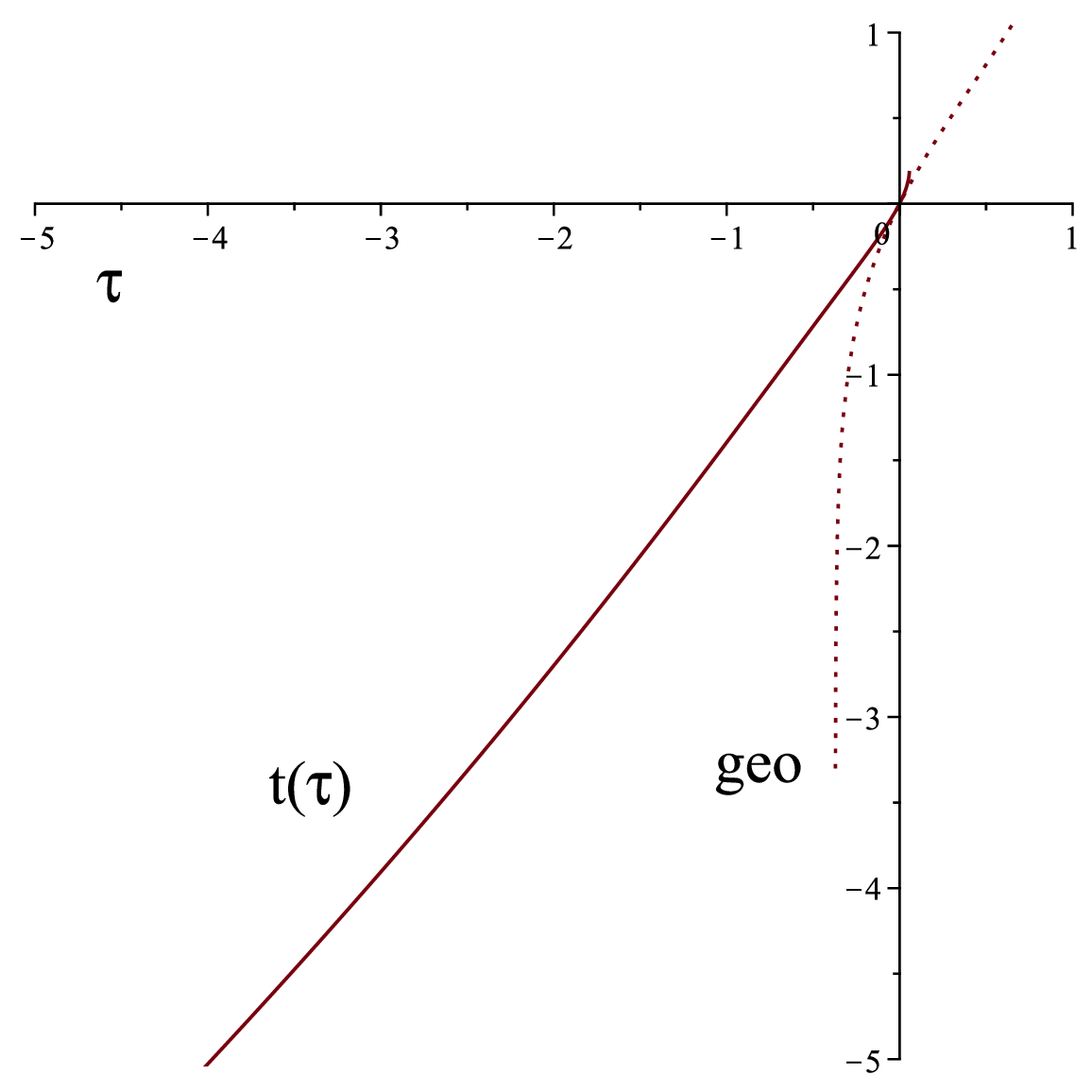}
\caption{\label{fig:t_di_tau_PR} Comparisons of temporal motions in presence of a 
PR external force and for a geodesic. Parameters and initial conditions are 
those of Fig. \ref{fig:rho_di_tau_PR}.
 }
\end{figure}

\begin{figure}
\includegraphics[scale=0.35]{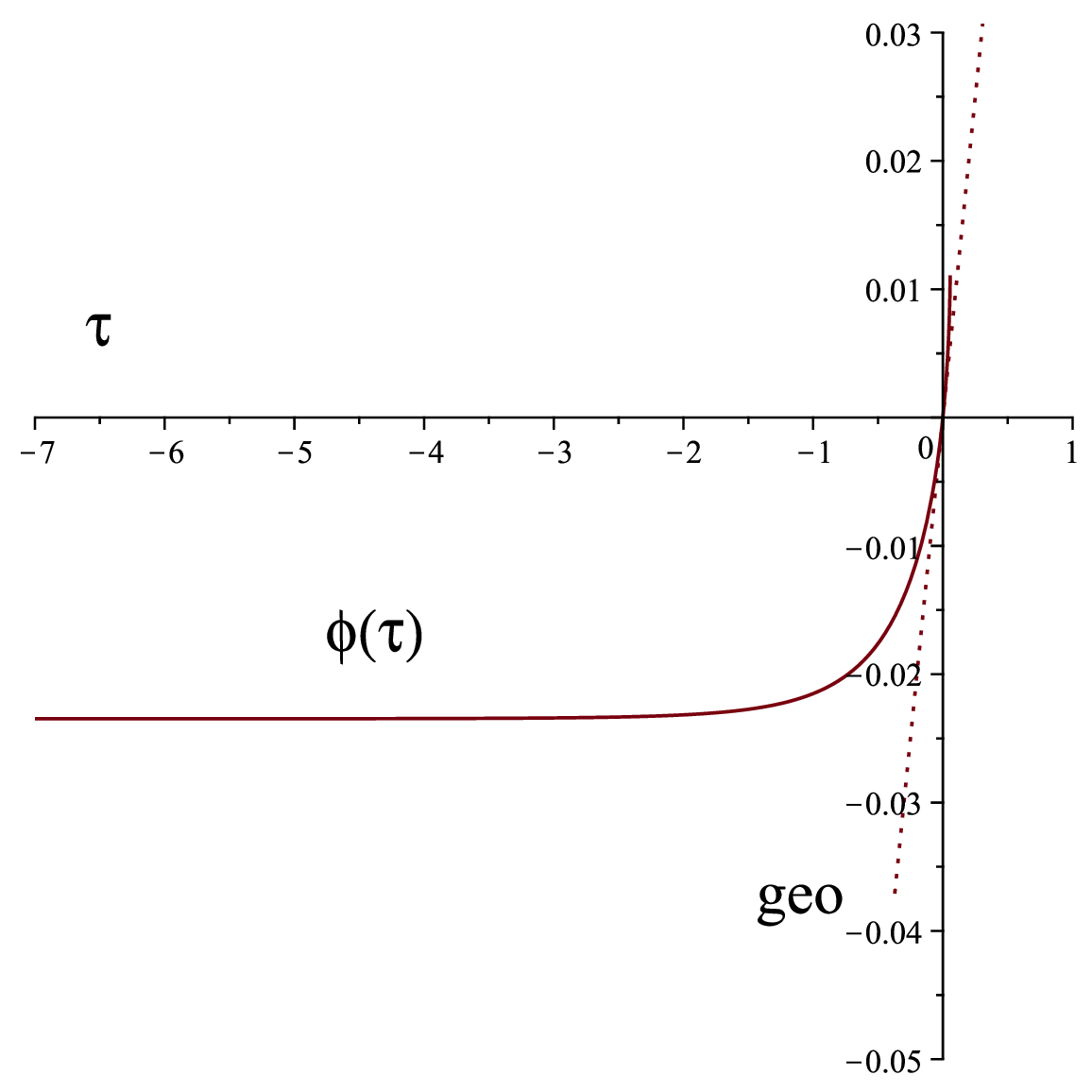}
\caption{\label{fig:phi_di_tau_PR} Comparisons of azimuthal motions in presence of a 
PR external force and for a geodesic. Parameters and initial conditions are 
those of Fig. \ref{fig:rho_di_tau_PR}.
 }
\end{figure}

\section{Can Nariai spacetime support a radiation field?}

Let us consider a photon field as described by the energy-momentum tensor
\begin{equation}
T^{\mu\nu}=\Phi_0 K^\mu K^\nu ,
\end{equation}
where $\Phi_0$ is a constant and
\begin{equation}
K=\frac{1}{(1-\rho^2)}\partial_T \pm \partial_\rho
\end{equation}
is a null (equatorial) geodesic of the background, inward ($-$) 
or outward ($+$) radially directed and
$T$ is divergence-free, $T^{\mu\nu}{}_{;\nu}=0$, and represents 
a photon field distribution on the background.

\begin{figure}
\includegraphics[scale=0.35]{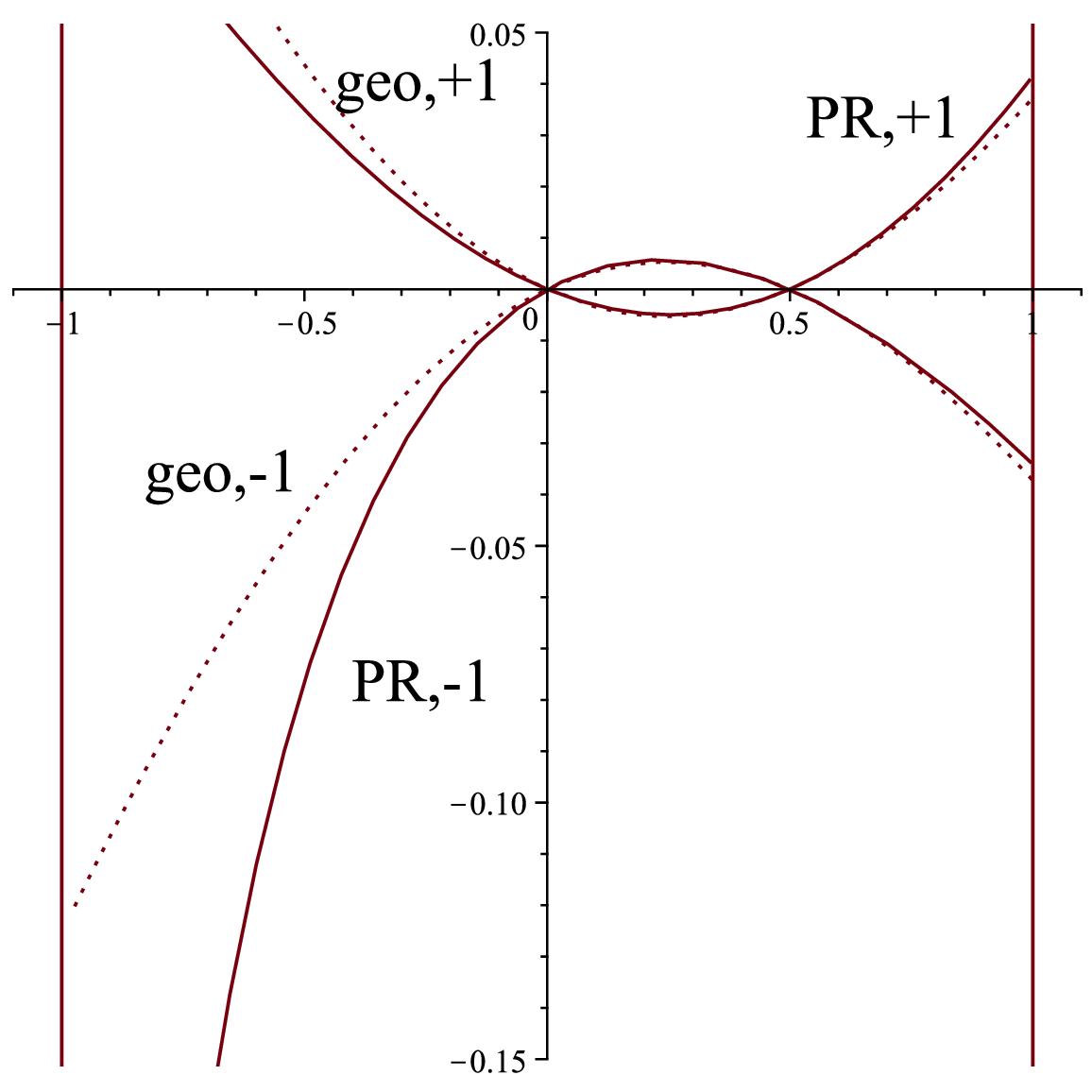}
\caption{\label{fig:light} Comparisons of motions in presence of a 
PR external force obtained with a superimposed radiation field 
and for a geodesic. On the $X$ axis we have $X=\rho(\tau)\cos(\phi(\tau))$ 
and on the $Y$ axis $Y=\rho(\tau)\sin(\phi(\tau))$. Parameters are 
$\Phi_0=1, \tilde \sigma=2$. The cases $\pm=1$ (integration radially 
outward, $+1$) and $\pm=-1$ (integration radially inward, $-1$) are highlighted. 
Initial conditions in all cases are chosen so that $t(0)=0$, $\phi(0)=0$, 
$\rho(0)=0.5$, $t'(0)=2$, $\phi'(0)=0.1$.
Dotted curves correspond to geodesics, $\tilde \sigma=0$, 
integrated radially inward or outward.
}
\end{figure}

Let $u$ be the four-velocity of a (massive) particle moving on the equatorial 
plane as before, Eq. \eqref{u_equat}, for which  $a(u)$ is given in Eq. \eqref{aU_compts}. 
The Poynting-Robertson-like force which the radiation field exerts on $u$, 
such that $m a(u)= F_{\rm ext}$, is then given by
\begin{equation}
F_{\rm ext}^\alpha=\sigma \Phi_0 (K\cdot u)[K^\alpha +(K\cdot u)u^\alpha].
\end{equation}
In terms of components 
\begin{eqnarray}
F_{\rm ext}^{T} &=& A [1-(u^t)^2(1-\rho^2)\pm u^t u^\rho ],
\nonumber\\
F_{\rm ext}^{\rho} &=& A[ -u^t u^\rho(1-\rho^2)\pm (1-\rho^2)\pm (u^\rho)^2 ] ,
\nonumber\\
F_{\rm ext}^{\phi} &=& A u^\phi [  -u^t(1-\rho^2)\pm u^\rho],
\end{eqnarray}
where
\begin{equation}
A=\frac{\sigma \Phi_0}{(1-\rho^2)^2}[-u^t(1-\rho^2)\pm u^\rho].
\end{equation}
To the first order in $\tilde \sigma$ the motion can be even studied analytically, 
in the sense that the $\phi$ equation decouples from the others and can be 
integrated exactly. The other two equations ($\rho$ and $T$) remain however 
too lengthy, and their study is better performed through plots.  
Examples of numerically integrated motions are illustrated in Fig. \ref{fig:light}.

\section{A test electromagnetic field in Nariai spacetime}

Let us consider a test electromagnetic field superimposed on Nariai spacetime.
In order to be fully general let us decompose the electromagnetic potential 
$A_\mu$ on a (1-form) basis of vector spherical harmonics, 
which in this case are given  by
\begin{eqnarray}
Z_1^\flat &=&Y_{lm}(\theta,\phi) dt 
\nonumber\\
Z_2^\flat&=&\frac{Y_{lm}(\theta,\phi)}{(1-\rho^2)}d\rho 
\nonumber\\
Z_3^\flat&=& \frac{\rho}{\sqrt{L}} \left[\partial_\theta 
Y_{lm}(\theta,\phi)  d\theta + i mY_{lm}(\theta,\phi)  d\phi\right]
\nonumber\\
&=&  \frac{\rho}{\sqrt{L}} \left[\partial_\theta Y_{lm}(\theta,\phi)  
d\theta + \partial_\phi Y_{lm}(\theta,\phi)  d\phi\right]
\nonumber\\
&=& \frac{\rho}{\sqrt{L}} d Y_{lm}(\theta,\phi)
\nonumber\\
Z_4^\flat&=& \frac{\rho}{\sqrt{L}} \left[\frac{1}{\sin\theta}
i mY_{lm}(\theta,\phi) d\theta \right.
\nonumber\\
&&\left. -\sin \theta \partial_\theta Y_{lm}(\theta,\phi) d\phi \right]
\nonumber\\
&=& \frac{\rho}{\sqrt{L}} \left[\frac{1}{\sin\theta}\partial_\phi 
Y_{lm}(\theta,\phi) d\theta \right.
\nonumber\\
&&\left. -\sin \theta \partial_\theta Y_{lm}(\theta,\phi) d\phi \right] ,
\end{eqnarray}
where $L=l(l+1)$
and we recall the relation
\begin{equation}
\partial_{\theta\theta }Y_{lm} +\cot(\theta)\partial_\theta Y_{lm} 
+\left(L-\frac{m^2}{\sin^2(\theta)}\right)Y_{lm}=0.
\end{equation}
The most general $A^\flat =A_\mu dx^\mu$ is then given by
\begin{eqnarray}
A^\flat&=& \sum_{lm}\int \frac{d\omega}{2\pi} \frac{e^{-i\omega t}}
{\rho}\Bigr[iZ_1^\flat u_1(\rho)+Z_2^\flat u_2(\rho)
\nonumber\\
&+& \frac{1}{\sqrt{L}}(Z_3^\flat u_3(\rho)+Z_4^\flat u_4(\rho))\Bigr].
\end{eqnarray}
The functions $u_i(\rho)$ are such that
\begin{eqnarray}
\frac{d^2 u_1}{d\rho^2}&=&-\left[\frac{A}{(1-\rho^2)^2}+\frac{2}{\rho^2}\right]u_1 
-\frac{2\rho \omega}{(1-\rho^2)^2}u_2+\frac{2}{\rho}\frac{d u_1}{d\rho},
\nonumber\\
\frac{d^2 u_4}{d\rho^2}&=& \frac{2\rho}{(1-\rho^2)}\frac{d u_4}{d\rho}
-\frac{A}{(1-\rho^2)^2}u_4,
\nonumber\\
\frac{du_3}{d\rho}&=& -\frac{\omega}{\rho(1-\rho^2)}\left(\frac{du_1}{d\rho}
-\frac{u_1}{\rho} \right)-\frac{A}{\rho(1-\rho^2)^2}u_2,
\end{eqnarray}
where $A=\omega^2-L(1-\rho^2)$.
These equations should be supplemented by the Lorenz gauge condition \cite{Lorenz}
\begin{equation}
\frac{du_2}{d\rho}-\frac{u_2}{\rho}=\frac{\omega}{(1-\rho^2)}u_1+\rho u_3.
\end{equation}
The variables $u_i$ turn out to be all coupled, except for $u_4$ which is 
formally solved in terms of Legendre functions
\begin{equation}
u_4(\rho) = C_1 P_{R,S}(\rho)+C_2 Q_{R,S}(\rho),
\end{equation}
where $C_1$ and $C_2$ are two integration constants and
\begin{equation}
R=\frac{\sqrt{1-4L}-1}{2} ,\qquad S=i\omega.
\end{equation}
In order to decouple these equations we first of all introduce the rescalings
\begin{equation}
u_1(\rho)=\rho U_1(\rho) ,\qquad u_2(\rho)=\rho U_2(\rho),
\end{equation}
motivated by the presence of terms of the type $\frac{du_1}{d\rho}-\frac{u_1}{\rho}$.
Upon defining 
\begin{equation}
f(\rho)=-\frac{((\rho^{2}-1)L + \omega^{2})}{(1-\rho^{2})^{2}}, \;
h(\rho)=-\frac{2 \omega \rho}{(1-\rho^{2})^{2}},
\label{(A9)}
\end{equation}
the new equations turn out to be
\begin{eqnarray}
\label{eqsUi}
\frac{d^2U_1}{d\rho^2}  &=& f(\rho)U_{1}(\rho)+h(\rho)U_{2}(\rho),
\nonumber\\
\frac{d U_2}{d\rho }   &=& -\frac{(1-\rho^2)}{2\rho}h(\rho)U_{1}(\rho)
+u_{3}(\rho),
\nonumber\\
\frac{d u_3}{d\rho }  &=& f(\rho)U_{2}(\rho)
+\frac{(1-\rho^2)}{2\rho} h(\rho) \frac{dU_1}{d \rho} .\qquad
\end{eqnarray}
From the second of these equation we isolate 
\begin{equation}
\label{eq_new_u3}
u_3(\rho)=\frac{d U_2}{d\rho} + \frac{(1-\rho^2)}{2\rho}h(\rho) U_1(\rho),  
\end{equation}
and we insert it into the third one, which becomes
\begin{equation}
\label{eqU2new}
\frac{d^2U_2}{d\rho^2} = -h(\rho)U_{1}(\rho)+f(\rho)U_{2}(\rho).
\end{equation}
At this stage, we focus on the coupled system obeyed by $U_1$ and $U_2$.
On defining the second-order differential operator
\begin{equation}
{\mathcal A}=\frac{d^2}{d\rho^2}-f(\rho),
\label{(10.13)}
\end{equation}
the coupled system just mentioned can be cast in the form
\begin{equation}
\label{coupled}
{\mathcal A}(U_1)=h U_2 ,\qquad {\mathcal A}(U_2)=-h U_1,
\end{equation}
or, equivalently,
\begin{equation}
\left(\begin{matrix}
{\mathcal A} & -h \cr h & {\mathcal A}
\end{matrix}\right)
\left(\begin{matrix}
U_{1} \cr U_{2}
\end{matrix}\right)=0.
\label{(10.14)}
\end{equation}
By introducing the notation
\begin{equation}
M_0=\left(\begin{matrix}
{\mathcal A} & -h \cr h & {\mathcal A}
\end{matrix}\right),\qquad
U=\left(\begin{matrix}
U_{1} \cr U_{2}
\end{matrix}\right)
\end{equation}
we obtain the concise form 
\begin{equation}
M_0 U=0.
\end{equation}

A repeated application of the operator ${\mathcal A}$ makes it possible  
to decouple these equations but leads to a fourth-order differential 
equation (more precisely, one obtains $U_1=-\frac{1}{h}{\mathcal A}(U_2)$ 
and subsequent insertion into the other equation, ${\mathcal A}(U_1)=h U_2$), 
i.e., it adds spurious solutions which should then be disentangled by taking 
into account the correct boundary conditions.
In order to avoid these spurious solutions one can proceed as follows.
We can diagonalize such a system by mapping its $2 \times 2$
matrix into (we are here relying upon Sec. 2 of Ref. \cite{EKMP})
\begin{equation}
M=\left(\begin{matrix}
1 & V \cr
W & 1
\end{matrix}\right)
\left(\begin{matrix}
{\mathcal A} & -h \cr h & {\mathcal A}
\end{matrix}\right)
\left(\begin{matrix}
1 & \alpha \cr \beta & 1
\end{matrix}\right),
\label{(10.15)}
\end{equation}
and then requiring that the off-diagonal differential operators $M_{12}$ and
$M_{21}$ should vanish for a suitable particular form of
$\alpha,\beta,V,W$. In the resulting calculation we bear in mind that,
by virtue of the Leibniz rule, one obtains (the $\cdot$ denotes here the
function acted upon by the derivative with respect to $\rho$)
\begin{equation}
\frac{d^2}{d\rho^2}(\alpha \cdot)=
\frac{d^{2}\alpha}{d\rho^{2}} (\cdot)
+\frac{d\alpha}{d\rho} \frac{d}{d\rho}(\cdot)
+\alpha \frac{d^2}{d\rho^2}(\cdot),
\label{(10.16)}
\end{equation}
jointly with an entirely analogous formula for $\beta$. In light of Eq.
\eqref{(10.15)}, the four operators of the matrix $M$ are the ones in
 Appendix \ref{appA}. On requiring that the off-diagonal operators should be
identically vanishing, we find therefore for $M_{12}$:
\begin{equation}
\alpha=-V,\qquad  \alpha'(\rho)=0 \qquad \forall \rho,
\label{(10.17)}
\end{equation}
\begin{equation}
h(V\alpha-1)=0 \Longrightarrow V^{2}+1=0,
\label{(10.18)}
\end{equation}
hence
\begin{equation}
V=\pm i, \; \alpha=\mp i.
\label{(10.19)}
\end{equation}
The analogous procedure for the vanishing of $M_{21}$ yields
\begin{equation}
W=\pm i, \; \beta=-W= \mp i.
\label{(10.20)}
\end{equation}
Eq. \eqref{(A1)} yields eventually
\begin{equation}
M_{11}=2 \frac{d^2}{d\rho^2}-2 (f(\rho)-ih(\rho)),
\label{(10.21)}
\end{equation}
while Eq. \eqref{(A4)} leads to
\begin{equation}
M_{22}=2 \frac{d^2}{d\rho^2}-2 (f(\rho)+ih(\rho)),
\label{(10.22)}
\end{equation}
that is
\begin{equation}
\frac{M_{11}}{2}={\mathcal A}+ih,\qquad
\frac{M_{22}}{2}={\mathcal A}-ih=\frac{M_{11}^*}{2}.
\end{equation}
Last, on reverting to the coupled equations \eqref{coupled} 
we notice that
\begin{equation}
{\mathcal A}(U_1+iU_2)=-ih(U_1+iU_2), 
\end{equation}
that is, the linear combination $Z=U_1+iU_2$ decouples the system,
and our decoupled differential equations are solved by
\begin{eqnarray}
Z(\rho) &=&  C_1\rho_{-}^{1+\frac{i\omega}{2}}\rho_{+}^{1-\frac{i\omega}
{2}}{}_2F_1([1+{\mathcal L}, 2-{\mathcal L}],[2-i\omega ],\rho_{+})
\nonumber\\
&+& C_2\rho_{-}^{1+\frac{i\omega}{2}}\rho_{+}^{\frac{i\omega}
{2}}{}_2F_1([i\omega+{\mathcal L}, 1+i\omega-{\mathcal L}],[i\omega],\rho_{+})
\nonumber\\
\end{eqnarray}
where $C_i$ are arbitrary constants and we have defined
\begin{equation}
{\mathcal L}=\frac12 +\frac12 (1-4 L)^{1/2},\qquad
\rho_\pm =\frac{(\rho\pm 1)}{2}.
\end{equation}
A second independent solution is then the complex conjugate $Z^*$.

\section{A test gravitational field in Nariai spacetime}

The gravitational perturbations of Nariai spacetime are rather involved, 
despite the underlying symmetries (cylindrical in the $T-\rho$ part of the 
metric and spherical in the $\theta-\phi$ part). This is mainly due to 
the lack of asymptotic flatness of the metric which adds to the fact that 
it is not a vacuum solution, thus enhancing the differences with 
respect to the similar treatment of other contexts, e.g., black holes.
We split the metric according to
\begin{equation}
g^{\rm pert}_{\alpha\beta}=g^{\rm N}_{\alpha\beta}+h_{\alpha\beta},
\end{equation}
and Fourier-expand the perturbation $h$ with respect to the Killing 
coordinates $T$ and $\phi$, i.e.,
\begin{equation}
h_{\alpha\beta}(T,\rho,\theta,\phi)=\int \frac{dm}{2\pi} e^{im\phi}\int 
\frac{d\omega}{2\pi}e^{-i\omega t} \hat h_{\alpha\beta}(\omega, \rho,\theta, m).
\end{equation}
By inserting these expressions into the perturbed Einstein equations, 
$G(g^{\rm N}_{\alpha\beta}+h_{\alpha\beta})-(g^{\rm N}_{\alpha\beta}
+h_{\alpha\beta})=0$ we obtain a lengthy set of coupled equations, listed below 
in Table \ref{grav_pert} (with an abuse of notation, i.e., $\hat h_{\alpha\beta}$ 
is still denoted by $h_{\alpha\beta}$, without the hat in order 
to simplify notation). 

If one does not perform any gauge choice (see Table \ref{grav_pert} 
for the resulting equations), special exact solutions 
can be found in this case as well, and are expressible in terms of 
Legendre associated functions. 
However, upon choosing a Regge-Wheeler gauge and 
decomposing the metric in tensor spherical harmonics 
(see e.g. Eqs. A5-A18 of Ref. \cite{Sago:2002fe} for notation and 
conventions in the Schwarzschild case, here \lq\lq adapted" 
to the Nariai solution by simply replacing 
$1-\frac{2M}{r}\to 1-\rho^2$ in the metric and $r\to \rho$ in harmonics), 
i.e., expressing all metric components in terms of the functions 
$h_0(\rho)$, $h_1(\rho)$, $H_0(\rho)$, $H_1(\rho)$, $H_2(\rho)$, one 
finds a set of (coupled) ordinary differential equations to be 
distinguished in the odd (or magnetic) sector 
\begin{eqnarray}
\label{sys_magn}
\frac{d^2h_0}{d\rho^2} &=& -i\omega \frac{d h_1}{d\rho } +\frac{L-2}
{(1-\rho^2)}h_0  ,\qquad (G_{02})
\nonumber\\
\frac{d h_0}{d\rho } &=& -i \frac{\omega^2+ (2-L)(1-\rho^2)}{\omega}h_1  ,
\qquad (G_{12})
\nonumber\\
\frac{d h_1}{d\rho } &=&  -\frac{i\omega}{(1-\rho^2)^2} 
h_0+\frac{2\rho}{(1-\rho^2)} h_1 ,\qquad (G_{22})\qquad
\end{eqnarray}
and the even (or electric) sector
\begin{eqnarray}
\label{sys_ele}
\frac{d^2K}{d\rho^2} &=&  \frac{L}{2\rho^2(1-\rho^2)} H_2  
+ \frac{ \rho^2 (L+6)-4 }{2\rho^2(1-\rho^2)}K 
\nonumber\\
&+&\frac{ 5\rho^2-4 }{\rho(1-\rho^2)} \frac{d K}{d\rho } ,\qquad (G_{00})
\nonumber\\
\frac{d K}{d\rho }  &=&  \frac{iL}{2\rho^2\omega} H_1
+\frac{ \rho^2-2 }{\rho(1-\rho^2)} K ,\qquad (G_{01})
\nonumber\\
\frac{d H_1}{d\rho } &=& -\frac{2\rho}{(1-\rho^2)} H_1 
-\frac{i\omega}{(1-\rho^2)} H_2 \nonumber\\
&-& \frac{\rho^2i\omega}{(1-\rho^2)} K ,\qquad (G_{02})
\nonumber\\
H_0 &=& H_1\frac{i\rho}{\omega}-\frac{\rho^2 [2(1+\omega^2)
-L(1-\rho^2)]}{L(1-\rho^2)} K ,\qquad (G_{11})
\nonumber\\
H_2  &=& H_0 ,\qquad (G_{22})
\nonumber\\
\end{eqnarray}
Notice that: 1) the replacements $H_1\to i \tilde H_1$ and $h_1\to 
i \tilde h_1$ (eventually $H_1\to i \omega \tilde H_1$ and 
$h_1\to i\omega \tilde h_1$) in this case make the system of equations 
real, and should be performed before attempting the decoupling; 
2) Replacing Eqs. $(G_{22})$ and $(G_{11})$ in Eq. $(G_{02})$ leads 
to the equivalent relation
\begin{equation}
\frac{d H_1}{d\rho }  =  \frac{3\rho}{(1-\rho^2)}  H_1 
+\frac{2i\rho^2\omega (1+\omega^2-L(1-\rho^2))}{L(1-\rho^2)^2} K, 
\end{equation}
which should be enough to obtain coupling with Eq. $(G_{01})$ in order to fully solve 
the even-parity perturbation system. Similarly, Eqs. $(G_{12})$ and $(G_{22})$ 
should be enough to fully solve the odd-parity system. Both cases lead to 
a coupled system of equations of the type
\beq
\begin{pmatrix}
\frac{dX}{d\rho}\\
\frac{dY}{d\rho}\\
\end{pmatrix}
=\begin{pmatrix}
C_{11}(\rho)& C_{12}(\rho)\\
C_{21}(\rho)& C_{22}(\rho)\\
\end{pmatrix}
\begin{pmatrix}
X\\
Y\\
\end{pmatrix}
\,.
\eeq
In fact, in the odd case we have $X=h_0$ and $Y=h_1$, while in the even one  
$X=K$ and $Y=H_1$. The equations can be easily decoupled. For example,
\bea
\frac{d^2h_0}{d\rho^2}  &=& -\frac{[(2-L)(1-\rho^2)+\omega^2]}{(1-\rho^2)^2}  h_0(\rho)\nonumber\\
&+& \frac{2 \omega^2\rho}{[(2-L)(1-\rho^2)+\omega^2](1-\rho^2)}\frac{dh_0}{d\rho}
\eea
A simple inspection of both the odd- and even-parity final set 
of equations seems to suggest that the solution for all metric functions 
are simpler (and solvable in terms of hypergeometric functions) with respect 
to the corresponding Schwarzschild case (which involve confluent Heun 
functions~\footnote{It is worth mentioning that a number of papers concerning 
Heun functions and their application to black hole and related spacetimes appeared 
recently in the literature, see e.g., \cite{Bautista:2023sdf,Aminov:2023jve,Arnaudo:2024rhv,Minucci:2024qrn}. 
All such studies will play a key role to obtain resummed expressions for several 
gauge-invariant quantities already computed by using a Post-Newtonian approximation.}).

In fact,
\beq
h_0(\rho)=(1-\rho^2)^{\frac{i\omega}{2}}\left[c_1 f_1(\rho)+c_2 f_2(\rho)\right]\,,
\eeq
where
\bea
f_1(\rho)&=&  \rho   \left[(\rho^2-1) (L-\omega^2+i\omega -2) h_{g_1}+i\omega h_{g_2}\right]\,,\nonumber\\
f_2(\rho)&=&  (\rho^2-1)\rho^2 \left( \frac{L-\omega^2}3+i\omega \right)h_{g_3}\nonumber\\ 
&+& [-1+(1+i\omega)\rho^2]  h_{g_4}\,, 
\eea
with
\bea
h_{g_1}&=& {}_2F_1 \left(\left[s_-+\frac54, s_++\frac54\right],\left[\frac32\right],\rho^2\right)\,,\nonumber\\
h_{g_2}&=& {}_2F_1 \left(\left[s_-+\frac14, s_++\frac14\right],\left[\frac12\right],\rho^2\right)\,, \nonumber\\
h_{g_3}&=& {}_2F_1 \left(\left[s_++\frac74, s_-+\frac74\right],\left[\frac52\right],\rho^2\right)\,,\nonumber\\
h_{g_4}&=& {}_2F_1 \left(\left[s_++\frac34, s_-+\frac34\right],\left[\frac32\right],\rho^2\right)\,,
\eea
having defined
\beq
s_\pm=\frac12 i\omega\pm \frac14 (9-4 L)^{1/2}\,.
\eeq
However, the complete analysis 
of the gravitational system (together with a generalization of the source terms to 
the case in which the perturbation is entirely due to a moving particle) 
will be conveniently postponed to future studies.

\begin{table*}  
\caption{\label{grav_pert} General set of coupled perturbation 
equations obtained by computing the Einstein equations for the metric 
$g^{\rm pert}_{\alpha\beta}=g^{\rm Nariai}_{\alpha\beta}+h_{\alpha\beta}$. 
The Killing variables $T$ and $\phi$ are ruled out by the Fourier transform. 
In the Table  the metric indices 
run from 1 through 4 and not from 0 through 3, so that for example 
$h_{14}$ corresponds to $h_{03}$. This (very convenient for practical purposes) 
re-labeling should not create confusion. Finally, an abbreviated notation 
here used is $[s,c]=[\sin\theta, \cos\theta]$.
}
\begin{ruledtabular}
\begin{tabular}{|l|l|}
Eq. 11 & $ \partial_{\rho\rho}h_{44} -m^2h_{22} +\frac{2\rho s c}{(1-\rho^2)} 
h_{23}+\frac{2i\rho m}{(1-\rho^2)} h_{24}-\frac{(m^2-2s^2)}{(1-\rho^2)} h_{33} 
+sc\partial_\theta h_{22}-2sc \partial_\theta h_{23}+\frac{2\rho s^2}
{(1-\rho^2)}\partial_\theta h_{23}-2im \partial_\rho h_{24}$\\
&$-\frac{\rho s^2}{(1-\rho^2)}\partial_\rho h_{33}-\frac{c s}{(1-\rho^2)}
\partial_\theta h_{33}-\frac{\rho}{(1-\rho^2)}\partial_\rho h_{44} 
-\frac{2c}{s(1-\rho^2)}\partial_\theta h_{44}+s^2\partial_{\theta\theta} 
h_{22}-2 s^2\partial_{\rho\theta}h_{23}+s^2\partial_{\rho\rho}h_{33}$\\
&$+\frac{1}{(1-\rho^2)}\partial_{\theta\theta}h_{44} +\frac{2}{(1-\rho^2)s^2} h_{44} =0  $\\

\hline 
Eq. 12 &$\partial_{\rho\theta}h_{13}+\frac{m^2}{s^2}h_{12}+\frac{2\rho c }
{ s (1-\rho^2)}h_{13}+i\omega \partial_\rho h_{33}+\frac{i\rho\omega}
{s^2 (1-\rho^2)} h_{44}+\frac{\omega m}{ s^2} h_{24} -i\omega \partial_\theta h_{23} 
+\frac{i\rho\omega}{ (1-\rho^2)} h_{33} -\frac{c}{s}\partial_\theta h_{12} 
+\frac{c}{s}\partial_\rho h_{13}$\\
&$+\frac{2\rho}{ (1-\rho^2) }\partial_\theta h_{13}  
+\frac{im}{s^2}\partial_\rho h_{14} +\frac{2i\rho m}{s^2 (1-\rho^2)}h_{14}
-\frac{i c\omega}{ s}h_{23}+\frac{i\omega}{s^2}\partial_\rho h_{44}
-\partial_{\theta\theta }h_{12}=0$\\

\hline 
Eq. 13 &$ \partial_{\rho\rho}h_{13}-\frac{(m^2-2s^2)}{ s^2 (1-\rho^2)} h_{13}
+\frac{ ic\omega}{s (1-\rho^2)} h_{33} +\frac{i \omega c}{s^3 (1-\rho^2)} 
h_{44} +\frac{2\rho}{ (1-\rho^2) }\partial_\theta h_{12}-\frac{i m}
{ s^2 (1-\rho^2)} \partial_\theta h_{14} -i\omega \partial_\theta h_{22} 
+i\omega \partial_{\rho}h_{23}$\\
&$ -\frac{i\omega}{ s^2 (1-\rho^2)}\partial_\theta h_{44}-\partial_{\rho\theta}h_{12}$\\

\hline
Eq. 14 &$ \partial_{\rho\rho}h_{14}+\frac{2i\rho m }{(1-\rho^2)}h_{12} 
+\frac{2}{(1-\rho^2)}h_{14}+\omega m h_{22}+\frac{m\omega}{(1-\rho^2)} h_{33} 
-i m\partial_\rho h_{12}-\frac{i m}{(1-\rho^2)}\partial_\theta h_{13} 
-\frac{c}{s(1-\rho^2)}\partial_\theta h_{14} +i\omega \partial_\rho h_{24} $\\
&$ +\frac{1}{(1-\rho^2)}\partial_{\theta\theta}h_{14} +\frac{i m c}{s(1-\rho^2)} h_{13}=0$\\

\hline
Eq. 22 &$\partial_{\theta \theta}h_{44}-\frac{2i\omega s^2}{(1-\rho^2)}
\partial_\theta h_{13}-\frac{sc}{(1-\rho^2)}\partial_\theta h_{11}+2\rho s^2 
\partial_\rho h_{33}-sc \partial_\theta h_{33}-\rho \partial_\rho h_{44}
-\frac{2c}{s}\partial_\theta h_{44}-\frac{s^2}{(1-\rho^2)}\partial_{\theta\theta}
h_{11}+\frac{m^2}{(1-\rho^2)} h_{11}$\\
&$ +\frac{2\omega m}{(1-\rho^2)} h_{14} +2 \rho s c h_{23}+2i\rho m h_{24} 
+\frac{[s^2\omega^2-(1-\rho^2) (m^2-2s^2)]}{(1-\rho^2)} h_{33} 
-\frac{(-\omega^2s^2-2(1-\rho^2))}{ s^2 (1-\rho^2)} h_{44}-\frac{2i\omega s c}
{(1-\rho^2)} h_{13} =0$\\

\hline
Eq. 23 &$\partial_{\rho\theta}h_{44} 
+ \frac{[s^2\omega^2-(1-\rho^2) (m^2-2s^2)]}{(1-\rho^2)}h_{23}
-\frac{\rho s^2}{(1-\rho^2)^2}\partial_\theta h_{11}  -\frac{i s^2\omega}
{(1-\rho^2)}\partial_\theta h_{12}  -\frac{i s^2\omega}{(1-\rho^2)}
\partial_\rho h_{13}+\rho s^2 \partial_\theta h_{22}$\\
&$-im \partial_\theta h_{24} -sc \partial_\rho h_{33} -\frac{c}{s}
\partial_\rho h_{44}-\frac{s^2}{(1-\rho^2)}\partial_{\rho\theta}h_{11}=0$\\

\hline
Eq. 24 &$\partial_{\theta\theta}h_{24} +\frac{\omega m}{(1-\rho^2)} h_{12}
+ i m\rho h_{22}+\frac{m ic}{s}h_{23} -h_{24}\frac{(2\rho^2-2-\omega^2)}
{(1-\rho^2)}-\frac{i m}{(1-\rho^2)} \partial_\rho h_{11}-\frac{i\omega}
{(1-\rho^2)}\partial_\rho h_{14}-im \partial_\theta h_{23} -\frac{c}{s}\partial_\theta h_{24}$\\
&$ +i m\partial_\rho h_{33} -\frac{i m\rho}{(1-\rho^2)^2}h_{11}=0$\\

\hline
Eq. 32 &$ \partial_{\rho\theta}h_{11} 
-\frac{[s^2\omega^2-(1-\rho^2) (m^2-2s^2)]}{s^2}h_{23}+\frac{\rho}{(1-\rho^2)}
\partial_{\theta}h_{11}+i\omega\partial_\theta h_{12} +i\omega\partial_\rho h_{13}  
- (1-\rho^2)\rho \partial_\theta h_{22}$\\
&$+\frac{i m(1-\rho^2)}{s^2}\partial_\theta h_{24} + \frac{(1-\rho^2) c}{s}
\partial_\rho h_{33}+\frac{(1-\rho^2) c}{s^3}\partial_\rho h_{44}
-\frac{(1-\rho^2)}{s^2}\partial_{\rho\theta}h_{44}=0$\\

\hline
Eq. 33 &$ \partial_{\rho\rho}h_{11} -\frac{(-2+m^2+2 c^2-m^2\rho^2)}
{(1-\rho^2)^2 s^2} h_{11}-\frac{2i\omega\rho}{(1-\rho^2)} h_{12}
-\frac{2\omega m}{ (1-\rho^2) s^2} h_{14}
-h_{22}
\frac{-m^2(1-\rho^2)+s^2(\omega^2+2-4\rho^2)}{s^2}
-\frac{4\rho c}{s} h_{23} $\\
&$ -\frac{4i m\rho}{s^2} h_{24}-\frac{\omega^2}{ (1-\rho^2) s^2} h_{44}
+\frac{\rho}{(1-\rho^2)} \partial_\rho h_{11} +\frac{c}{s(1-\rho^2)}\partial_\theta h_{11}  
+\frac{2im (1-\rho^2)}{s^2} \partial_\rho h_{24} -(1-\rho^2)\rho\partial_\rho h_{22} 
-\frac{c(1-\rho^2)}{s}\partial_\theta h_{22}$\\
&$+\frac{2c}{s(1-\rho^2)}\partial_\rho h_{23}+2i\omega \partial_\rho h_{12} 
+2\partial_\rho h_{44} \frac{\rho}{s^2}
-\frac{(1-\rho^2)}{s^2}\partial_{\rho\rho}h_{44}+\frac{2i\omega c}{s(1-\rho^2)} h_{13} =0$\\

\hline
Eq. 34 &$  \partial_{\rho\theta}h_{24} -\frac{i m c}{s(1-\rho^2)^2} h_{11}
-\frac{\omega m}{(1-\rho^2)^2} h_{13}+\frac{icm}{s} h_{22}
-\frac{2i\rho m}{(1-\rho^2)} h_{23}+\frac{4\rho c}{s(1-\rho^2)} h_{24}
+  \frac{im}{(1-\rho^2)^2}\partial_\theta h_{11}+\frac{i\omega}{(1-\rho^2)^2} 
\partial_\theta h_{14}$\\
&$-i m \partial_\theta h_{22}+im\partial_\rho h_{23}
-\frac{2c}{s}\partial_\rho h_{24}-\frac{2\rho}{(1-\rho^2)}\partial_\theta 
h_{24}-\frac{2i\omega c}{s(1-\rho^2)^2} h_{14}=0$\\

\hline
Eq. 44 &$ \partial_{\theta\theta}h_{11}-2i\omega \rho h_{12}+ (1-\rho^2)
(4\rho^2-\omega^2-2)h_{22}-\omega^2h_{33}+\rho\partial_\rho h_{11}+2i 
(1-\rho^2)\omega\partial_\rho h_{12}+2i\omega\partial_\theta h_{13} 
- (1-\rho^2)^2\rho \partial_\rho h_{22}$\\
&$-4\rho (1-\rho^2)\partial_\theta h_{23}+2\rho (1-\rho^2)\partial_\rho h_{33}
+(1-\rho^2) \partial_{\rho\rho}h_{11}
-(1-\rho^2)^2\partial_{\theta\theta} h_{22}  +2 (1-\rho^2)^2\partial_{\rho\theta}
h_{23}- (1-\rho^2)^2 \partial_{\rho\rho}h_{33}$\\
&$+\frac{2}{(1-\rho^2)} h_{11}=0$\\

\hline
Comp. &$  
\partial_{\rho\rho}h_{33}-\frac{m^2}{ s^2 (1-\rho^2)^2} h_{11}-\frac{2\omega m}
{s^2 (1-\rho^2)^2} h_{14}-\frac{2c\rho}{s(1-\rho^2)} h_{23}
-\frac{2i \rho m }{s^2(1-\rho^2)}h_{24}
-\frac{\omega^2}{s^2(1-\rho^2)^2} h_{44}+\frac{c}{s(1-\rho^2)^2}\partial_\theta 
h_{11}+\frac{2\rho}{(1-\rho^2)}\partial_\theta h_{23}$\\
&$ -\frac{\rho}{(1-\rho^2)}\partial_\rho h_{33}
+\frac{\rho}{s^2(1-\rho^2)}\partial_\rho h_{44} +\partial_{\theta\theta}h_{22} 
-2\partial_{\rho\theta} h_{23} +\frac{2i\omega c}{s(1-\rho^2)^2} h_{13}=0$\\
 \end{tabular}
\end{ruledtabular}
\end{table*}

\section{Concluding remarks}

Our work on Nariai spacetime was initiated in Ref. \cite{BE2024}, but here
we have focused on a completely different set of properties, paying 
first attention to some curvature 
aspects, connecting for example the sectional curvatures of a timelike 2-section 
and of a spacelike 2-section to the spacetime curvature itself by analyzing 
the Kretschmann invariant properties. 
We have also studied both geodesic motion (obtaining a fully original, complete analytic 
integration of the orbits) and accelerated one (in particular, the accelerated 
motion of an observer at rest with respect to the spherical coordinates, 
and in radial motion with respect to them). 

Moreover, by exploiting the algebraic speciality of Nariai spacetime (which is of Petrov 
type D) we have investigated perturbations due to a  
scalar field (treated in a fully analytical way). In the 
latter case, our results will be relevant as (exact type) cosmological perturbations 
if Nariai spacetime is assumed as a cosmological model. We will leave this 
problem to future investigations and projects. Indeed, it is 
hard to investigate global solutions of the equations
obeyed by Fourier modes even just for a scalar field (see e.g. Refs. 
\cite{Per1,Per2} for the analogous situation in a Schwarzschild spacetime),
and the resulting integral representation of the scalar field is
technically hard as well.  

Furthermore, we have studied conditions that make it possible for Nariai spacetime to support 
a superimposed test fluid in the special situation of a rest frame of the fluid 
radially moving with respect to cylindrical-like form of the background metric. 
We have displayed two examples of radially moving observers taken as identifying 
the rest frame of the fluid: an observer with constant frame components and another 
with anisotropic radial velocity. We have then discussed how a test particle moving 
in Nariai spacetime with a fluid superimposed can feel the presence of the fluid 
and can be deviated from the original (geodesic or nongeodesic) motion.

Eventually, we have developed original calculations in order to study
the possible occurrence of test electromagnetic or gravitational fields in
Nariai spacetime. In the electromagnetic case, the imposition of a gauge
condition and the exploitation of a decoupling technique have led to a
full analytic solution, whereas the gravitational case remains in general rather involved. 
The Regge-Wheeler gauge, however, leads to additional simplifications.

Another technical issue for future research is the actual analytic or
numerical integration of Fourier modes in order to evaluate a massive
scalar field in Nariai spacetime. Our exact results in $\omega$ space will 
be useful as soon as one will be able to provide testbeds for numerical results 
when these will become available.

\section*{Acknowledgements}
D.B. thanks A. Geralico and B. Wardell for informative discussions.
D.B. and G.E. are grateful to INDAM for membership. 

\section*{Data Availability} 
No data were created or analyzed in this study.

\appendix
\section{The matrix of differential operators in Sec. X}
\label{appA}

By virtue of Eq. \eqref{(10.15)}, the matrix $M$ of differential
operators has elements
\begin{eqnarray}
\label{(A1)}
M_{11}&=&(1+V \beta)\frac{d^2}{d \rho^2}
+V \beta'(\rho)\frac{d}{d \rho}\nonumber\\
&+&V \beta''(\rho)
-(1+V \beta)f+h(V-\beta),
\\
\label{(A2)}
M_{12}&=&(\alpha+V)\frac{d^2}{d\rho^2}
+\alpha'(\rho)\frac{d}{d\rho}\nonumber\\
&+&\alpha''(\rho)
-(\alpha+V)f+h(V \alpha-1), 
\end{eqnarray}
\begin{eqnarray}
\label{(A3)}
M_{21}&=&(\beta+W)\frac{d^2}{d\rho^2}
+\beta'(\rho)\frac{d}{d\rho}\nonumber\\
&+&\beta''(\rho)
-(\beta+W)f+h(1-W \beta),
\\
\label{(A4)}
M_{22}&=&(1+W \alpha)\frac{d^2}{d \rho^2}
+W \alpha'(\rho)\frac{d}{d \rho}\nonumber\\
&+&W \alpha''(\rho)
-(1+W \alpha)f+h(\alpha-W).
\end{eqnarray}

\section{Killing equations in Nariai spacetime}
\label{appB}

Let us consider the dS-like form of the Nariai spacetime metric and a generic vector
\begin{equation}
\xi=\xi^T\partial_T+\xi^\rho\partial_\rho+\xi^\theta\partial_\theta+\xi^\phi\partial_\phi .
\end{equation}
The Killing equations 
\begin{equation}
\nabla_{(\alpha}\xi_{\beta)}=0,
\end{equation}
imply the following set of coupled equations
\begin{eqnarray}
\partial_T \xi^T&=&\frac{\rho}{(1-\rho^2)}\xi^\rho,
\nonumber\\
\partial_\rho \xi^T&=& \frac{1}{(1-\rho^2)^2}\partial_T \xi^\rho,
\nonumber\\
\partial_\theta \xi^T&=& \frac{1}{(1-\rho^2)}\partial_T \xi^\theta,
\nonumber\\
\partial_\phi \xi^T&=& \frac{\sin^2\theta}{(1-\rho^2)}\partial_T \xi^\phi,
\end{eqnarray}
\begin{eqnarray}
\partial_\rho \xi^\rho&=&-\frac{\rho}{(1-\rho^2)}\xi^\rho, 
\nonumber\\
\partial_\rho \xi^\theta&=&-\frac{1}{(1-\rho^2)}\partial_\theta \xi^\rho, 
\nonumber\\
\partial_\rho \xi^\phi&=&-\frac{1}{(1-\rho^2)\sin^2\theta}\partial_\phi \xi^\rho,
\nonumber\\
\partial_\theta \xi^\theta&=&0,
\end{eqnarray}
and
\begin{eqnarray}
\partial_\theta \xi^\phi&=&-\frac{1}{\sin^2\theta}\partial_\phi \xi^\theta,
\nonumber\\
\partial_\phi \xi^\phi&=&-\cot \theta \xi^\theta.
\end{eqnarray}
Trivial solutions to these equations are $\xi=\partial_T$ and $\xi=\partial_\phi$. 
Other solutions exist which correspond to the two separate sectors, 
$(T,\rho)$ and $(\theta,\phi)$. For example, 
\begin{eqnarray}
\xi&=&\frac{\rho}{(1-\rho^2)^{1/2}}\left(C_1e^T-C_2e^{-T}\right)\partial_T 
\nonumber\\
&+&(1-\rho^2)^{1/2}\left(C_1 e^T+C_2e^{-T}\right)\partial_\rho,
\end{eqnarray}
is a Killing vector of the $(T,\rho)$ sector for any choice of the constants $C_1$ and $C_2$.
Similarly, in the $(\theta,\phi)$ sector
\begin{eqnarray}
\xi&=&(c_1\sin(\phi)+c_2\cos(\phi))\partial_\theta  
\nonumber\\
&+& \cot\theta (c_1\cos(\phi)-c_2\sin(\phi))\partial_\phi,
\end{eqnarray}
is a Killing vector for any choice of the constants $c_1$ and $c_2$.

\end{document}